\documentclass[10pt,journal,compsoc]{IEEEtran}

\usepackage{caption}
\usepackage{hyperref}
\usepackage{subfig}

\ifCLASSOPTIONcompsoc
  \usepackage[nocompress]{cite}
\else
  \usepackage{cite}
\fi
\ifCLASSINFOpdf
\else
\fi
\usepackage{etex}

\usepackage{booktabs}
\usepackage{pgfplotstable}

\usepackage{adjustbox}

\usepackage[T1]{fontenc}

\usepackage{amsmath}
\usepackage{amsthm}
\usepackage{amssymb}

\usepackage{comment}

\usepackage{array}
\usepackage{tabu}   % Wide lines in tables
\usepackage{xspace} % Non-eatable spaces in macros
\usepackage{graphicx}
\graphicspath{{figures/}}
\usepackage{hyperref}
\usepackage[all]{hypcap}
\usepackage{color}
\usepackage{xcolor}
\definecolor{mauve}{rgb}{0.58,0,0.82}

\let\chapter\section
\usepackage[ruled, vlined, linesnumbered]{algorithm2e}

\SetKw{True}{true}
\SetKw{False}{false}
\SetKwData{typeInt}{Int}
\SetKwData{typeRat}{Rat}
\SetKwData{Defined}{Defined}
\SetKwFunction{parseStatement}{parseStatement}

\usepackage{todonotes}

\usepackage{listings}
\usepackage{tikz}
\usepackage{tikz-qtree}
\usepackage{fixltx2e}
\usepackage{pdflscape}
\usepackage{afterpage}
\usepackage{capt-of}% or use the larger `caption` package

\usepackage{lipsum}

\usepackage{multirow}
\usepackage{colortbl}

\begin{document}
\title{Automated Discovery of Process Models from Event Logs: Review and Benchmark}
\author{Adriano Augusto, Raffaele Conforti, Marlon~Dumas, Marcello~La Rosa, \\Fabrizio Maria Maggi, Andrea Marrella, Massimo Mecella, Allar Soo
\IEEEcompsocitemizethanks{
\IEEEcompsocthanksitem  A. Augusto, M. Dumas, F.M. Maggi, A. Soo are with the University of Tartu, Estonia.\protect\\
E-mail: \{adriano.augusto,marlon.dumas,f.m.maggi\}@ut.ee,\protect\\ allar.soo@gmail.com
\IEEEcompsocthanksitem A. Augusto, R. Conforti, M. La Rosa are with Queensland University of Technology, Australia.\protect\\
E-mail: \{a.augusto,raffaele.conforti,m.larosa\}@qut.edu.au
\IEEEcompsocthanksitem A. Marrella, M. Mecella are with Sapienza Universit\'a di Roma, Italy.\protect\\
E-mail: \{marrella,mecella\}@diag.uniroma1.it
}
\thanks{Manuscript received XX; revised XX.}
}

\IEEEtitleabstractindextext{%
\begin{abstract}
Process mining allows analysts to exploit logs of historical executions of business processes to extract insights regarding the actual performance of these processes. One of the most widely studied process mining operations is automated process discovery. An automated process discovery method takes as input an event log, and produces as output a business process model that captures the control-flow relations between tasks that are observed in or implied by the event log. Various automated process discovery methods have been proposed in the past two decades, striking different tradeoffs between scalability, accuracy and complexity of the resulting models. However, these methods have been evaluated in an ad-hoc manner, employing different datasets, experimental setups, evaluation measures and baselines, often leading to incomparable conclusions and sometimes unreproducible results due to the use of closed datasets. This article provides a systematic review and comparative evaluation of automated process discovery methods, using an open-source benchmark and covering twelve publicly-available real-life event logs, twelve proprietary real-life event logs, and nine quality metrics. The results highlight gaps and unexplored tradeoffs in the field, including the lack of scalability of some methods and a strong divergence in their performance with respect to the different quality metrics used. 
\end{abstract}

}

\maketitle

\IEEEdisplaynontitleabstractindextext

% \fi
\IEEEpeerreviewmaketitle

% !TEX root = main.tex
\section{Introduction}\label{sec:introduction}

%  file is ....

Modern information systems maintain detailed trails of the business processes they support, including records of key process execution events, such as the creation of a case or the execution of a task within an ongoing case.
Process mining techniques allow analysts to extract insights about the actual performance of a process from collections of such event records, also known as \emph{event logs}~\cite{DBLP:journals/tkde/AalstWM04}. In this context, an event log consists of a set of traces, each trace itself consisting of the sequence of events related to a given case.

%Process mining methods allow analysts to exploit logs of historical executions of business processes in order to extract insights regarding the actual performance of these processes.
One of the most widely investigated process mining operations is \emph{automated process discovery}.
An automated process discovery method takes as input an event log, and produces as output a \emph{business process model} that captures the control-flow relations between tasks that are observed in or implied by the event log.

%Among other things, process mining techniques allow analysts to automatically discover a process model from an event log.
In order to be useful, such automatically discovered process models must accurately reflect the behavior recorded in or implied by the log.
Specifically, the process model discovered from an event log should be able to: \emph{(i)} generate each trace in the log, or for each trace in the log, generate a trace that is similar to it; \emph{(ii)} generate traces that are not in the log but that are identical or similar to traces of the process that produced the log; and \emph{(iii)} not generate other traces~\cite{ProcessMiningBook}. The first property is called \emph{fitness}, the second \emph{generalization} and the third \emph{precision}.
In addition, the discovered process model should be as simple as possible, a property that is usually quantified via \emph{complexity} measures.

The problem of automated discovery of process models from event logs has been intensively researched in the past two decades. Despite a rich set of proposals,
state-of-the-art automated process discovery methods suffer from two recurrent deficiencies when applied to real-life logs~\cite{DeWeerdt}: \emph{(i)} they produce large and spaghetti-like models; and \emph{(ii)} they produce models that either poorly fit the event log (low fitness) or over-generalize it (low precision or low generalization). Striking a tradeoff between these quality dimensions in a robust manner has proved to be a difficult problem.

%Several dozen automated process discovery methods have been proposed in the past two decades, striking different tradeoffs between scalability, accuracy and complexity of the resulting models.
%\todo[inline]{following paragraphs need revision}
%Despite two decades of intensive research leading to several dozen automated process discovery techniques, striking a tradeoff between the above four quality dimensions (fitness, precision, generalization and complexity) has proved elusive. When applied to real-life logs, the vast majority of automated process discovery methods (e.g., the Heuristics Miner~\cite{DBLP:conf/cidm/WeijtersR11} and its derivatives) produce large, spaghetti-like and oftentimes behaviorally incorrect (e.g., deadlocking) process models. Another state-of-the-art method, namely the Inductive Miner~\cite{DBLP:conf/apn/LeemansFA13}, produces structured and behaviorally correct process models with high fitness but very poor precision and generalization -- i.e., the resulting models grossly over-generalize the behavior observed in the log.

So far, automated process discovery methods have been evaluated in an ad hoc manner, with different authors employing different datasets, experimental setups, evaluation measures and baselines, often leading to incomparable conclusions and sometimes unreproducible results due to the use of non-publicly available datasets.
This work aims at filling this gap by providing: \emph{(i)} a systematic review of automated process discovery methods; and \emph{(ii)} a comparative evaluation of seven implementations of representative methods, using an open-source benchmark framework and covering twelve publicly-available real-life event logs, twelve proprietary real-life event logs, and nine quality metrics covering all four dimensions mentioned above (fitness, precision, generalization and complexity), as well as execution time.

%The review and evaluation results highlight gaps and unexplored tradeoffs in the field, including the lack of scalability of several proposals in the field and a strong divergence in the performance of different methods with respect to different quality metrics.

The outcomes of this research are a classified inventory of automated process discovery methods and a benchmark designed to enable researchers to empirically compare new automated process discovery methods against existing ones in a unified setting. The benchmark is provided as an open-source command-line Java application to enable researchers to replicate the reported experiments with minimal configuration effort.
%and to add their own process discovery methods.

The rest of the article is structured as follows. Section~\ref{sec:methodology} describes the search protocol used for the systematic literature review, whereas Section~\ref{sec:taxonomy} classifies the methods identified in the review. Then, Section~\ref{sec:benchmark} introduces the experimental benchmark and results, whereas Section~\ref{sec:discussion} discusses the overall findings and Section~\ref{sec:threats_to_validity} acknowledges the threats to the validity of the study. Finally,  Section~\ref{sec:related_work} relates this work to previous reviews and comparative studies in the field and Section~\ref{sec:conclusion} concludes the paper and outlines future work directions.

%\subsection{Problem Statement and Scope}
%\label{subsec:problem_statement}
%\subsection{Contribution}
%\label{subsec:contribution} 
% !TEX root = main.tex
\section{Search protocol}
\label{sec:methodology}

In order to identify and classify research in the area of automated process discovery, we conducted a \emph{Systematic Literature Review} (SLR)
through a scientific, rigorous and replicable approach as specified by Kitchenham~\cite{Kitchenham}.

First, we formulated a set of research questions to scope the search, and developed a list of search strings.
Next, we ran the search strings on different data sources.
Finally, we applied inclusion criteria to select the studies retrieved through the search.

 % (cf. Section \ref{subsec:research_question}), after which research strings were created (cf. Section \ref{subsec:search_string}) and data sources were selected (cf. Section \ref{subsec:data_source_selection}). Then, we identified inclusion and exclusion criteria (cf. Section \ref{subsec:inclusion_exclusion_criteria}) and methods for quality assessment (cf. Section \ref{subsec:quality_assessment}). Finally, we performed research studies selection (cf. Section \ref{subsec:study selection}), data extraction and analysis (cf. Section \ref{subsec:data_extraction_analysis})

\subsection{Research questions}
\label{subsec:research_question}

The objective of our SLR is to analyse research studies related to automated (business) process discovery from event logs.
This means, for example, that methods performing only trace clustering or log-filtering are not considered in our analysis.
To this aim, we formulated the following research questions:
\begin{itemize}
	\item[RQ1] What methods exist for \emph{automated} (business) process discovery from \emph{event logs}?
	\item[RQ2] What \emph{type of process models} can be discovered by these methods, and in which \emph{modeling language}?
	\item[RQ3] Which \emph{semantic} can be captured by a model discovered by these methods?
	\item[RQ4] What \emph{tools} are available to support these methods?
	\item[RQ5] What \emph{type of data} has been used to evaluate these methods, and from which application domains?
\end{itemize}

RQ1 is the core research question, which aims at identifying existing methods to perform  (business) process discovery from event logs. The other questions allow us to identify a set of classification criteria. Specifically, RQ2 categorizes the output of a method on the basis of the type of process model discovered (i.e., procedural, declarative or hybrid), and the specific modeling language employed (e.g., Petri nets, BPMN, Declare). RQ3 delves into the specific semantic constructs supported by a method (e.g., exclusive choice, parallelism, loops). RQ4 explores what tool support the different methods have, while RQ5 investigates how the methods have been evaluated and in which application domains.

\subsection{Search string development and validation}
\label{subsec:search_string}

Next, we developed four search strings by deriving keywords from our knowledge of the subject matter.
We first determined that the term ``process discovery''
is a very generic term which would allow us to retrieve the majority of methods in this area.
Furthermore, we used ``learning'' and ``workflow'' as synonyms of ``discovery'' and ``process'' (respectively).
This led to the following four search strings: (i) ``process discovery'', (ii) ``workflow discovery'', (iii) ``process learning'',
(iv) ``workflow learning''.
We intentionally excluded the terms ``automated'' and ``automating'' in the search strings, because these terms are often not explicitly used.

However, this led to retrieving many more studies than those that actually focus on automated process discovery, e.g., studies on process discovery via workshops or interviews. Thus, if a query on a specific data source returned more than one thousand results, we refined it by combining the selected search string with the term ``business'' or ``process mining'' to obtain more focused results, e.g., ``process discovery AND process mining'' or ``process discovery AND business''. According to this criterion, the final search strings used for our search were the following:

\begin{itemize}
\item[i.] ``process discovery AND process mining''
\item[ii.] ``process learning AND process mining''
\item[iii.] ``workflow discovery''
\item[iv.] ``workflow learning''
\end{itemize}

First, we applied each of the four search strings to Google Scholar, retrieving studies based on the occurrence of one of the search strings in the title, the keywords or the abstract of a paper. %This led to the identification of 2,820 studies.
Then, we used the following six popular academic databases: Scopus, Web of Science, IEEE Xplore, ACM Digital Library, SpringerLink, ScienceDirect, to double check the studies retrieved from Google Scholar.
%
%First, we applied each of the four search strings to Google Scholar, retrieving any study whose full text contained at least one of the search strings. This led to the identification of 2,820 studies.
%Successively, we used the following six popular academic databases: Scopus, Web of Science, IEEE Xplore, ACM Digital Library, SpringerLink, ScienceDirect, and retrieved studies based on the occurrence of one of the search strings in the title, the keywords or the abstract of a paper. This second search was meant to be a double check of the studies retrieved from Google Scholar.
%
We noted that this additional search did not return any relevant study that was not already discovered in our primary search. The search was completed in December 2017.

%Kitchenham \cite{Kitchenham} recommends to validate trial search strings against lists of already known primary studies. Accordingly, we examined an existing survey in the field of automated process discovery, namely \cite{DeWeerdt}, and verified that all publications contemplated in this survey were also found by our search. Given that \cite{DeWeerdt} refers to studies published prior to 2012, our search returned a large number of publications not covered in \cite{DeWeerdt}.

\subsection{Study selection}
\label{subsec:study selection}

As a last step, as suggested by \cite{Fink10,Okoli10,Randolph09,Torraco05}, we defined inclusion criteria to ensure an unbiased selection of relevant studies.
To be retained, a study must satisfy all the following inclusion criteria.
%The development of these criteria, as recommended in \cite{Kitchenham}, was based on the objective and the scope of this survey, as defined by the research questions distilled in Section~\ref{subsec:research_question}. This led to the following criteria.

\noindent
\begin{itemize}
\item[IN1] \emph{The study proposes a method for automated (business) process discovery from event logs}.
This criterion draws the borders of our search scope and it is direct consequence of RQ1.
\item[IN2] \emph{The study proposes a method that has been implemented and evaluated}.
This criterion let us exclude methods whose properties have not been evaluated nor analyzed.
\item[IN3] \emph{The study is published in 2011 or later}.
Earlier studies have been reviewed and evaluated by De Weerdt et al.\cite{DeWeerdt},
therefore, we decided to focus only on the successive studies.
Nevertheless, we performed a mapping of the studies assessed in 2011~\cite{DeWeerdt}
and their successors (when applicable), cf. Table~\ref{tab:superseded}.
\item[IN4] \emph{The study is peer-reviewed}.
This criterion guarantees a minimum reasonable quality of the studies included in this SLR.
\item[IN5] \emph{The study is written in English}.
\end{itemize}

\begin{table}[h!]
\centering
{\scriptsize{
\begin{tabular}{c|c}
	\hline
	$\alpha$, $\alpha^+$, $\alpha^{++}$~\cite{van2004workflow, alves2004process, wen2007mining} &
	$\alpha$\$~\cite{guo2015mining}
	\\
	AGNEs Miner~\cite{goedertier2009robust} & ---
	\\
	(DT) Genetic Miner~\cite{de2005genetic, de2007genetic} &
	Evolutionary Tree Miner~\cite{buijs2014quality}
	\\
	Heuristics Miner~\cite{weijters2003rediscovering, weijters2006process} &
	Heuristics Miner~\cite{augusto2017automated,mannhardt2017data,weijters2011flexible,de2014improving}
	\\
	ILP Miner~\cite{van2009process} &
	Hybrid ILP Miner~\cite{van2017discovering}	
\\\hline
\end{tabular}
}}
\caption{Methods assessed by De Weerdt et al.~\cite{DeWeerdt} (left) and the respective successors (right).}\label{tab:superseded}
\end{table}

Inclusion criteria IN3, IN4 and IN5 were automatically applied through the configuration of the search engines.
After the application of the latter three inclusion criteria, we obtained a total of 2,820 studies.
Then, we skimmed title and abstract of these studies to exclude those studies that were clearly not compliant with IN1.
As a result of this first iteration, we obtained 344 studies.

Then, we assessed each of the 344 studies against the inclusion criteria IN1 and IN2.
The (combined) assessment of IN1 and IN2 on the 344 selected studies was performed independently by two authors of this paper, whose decisions were compared in order to resolve inconsistencies with the mediation of a third author, when needed.
The assessment of IN1 was based on the accurate reading of the abstract, introduction and conclusion.
Whilst, to determine whether a study fulfilled IN2, we relied on the findings reported in the evaluation of the studies.
As result of the iterations, we found 86 studies matching the five inclusion criteria.

However, many of these studies refer to the same automated process discovery method,
i.e., some studies are either extensions, optimization, preliminaries or generalization of another study.
For such reason, we decided to group the studies by either the last version or the most general one.
When in doubt, the grouping decision was taken after a consultation with the main author. At the end of this process, as shown in Table~\ref{tab:SumTabel}, 35 main groups of discovery algorithms were identified.

The excel sheet available at \url{https://drive.google.com/open?id=1fW8WLXSwI2ntiPu3XVgsDI1cJUJr762G}
reports the 344 studies found after the first iteration.
For each of these studies, we explicitly refer to the inclusion criterion fulfilled for the study to be selected.
Furthermore, each selected study has a reference to the group it belongs to (unless it is the main study).

\begin{figure}[h!]
\centering
\resizebox {\columnwidth} {!} {
\begin{tikzpicture}
\begin{axis}[
    title={},
    grid=major,
    xlabel={Publication Year},
	x tick label style={/pgf/number format/1000 sep=},
    ylabel={\# Studies},
    xmin=2010, xmax=2018,
    ymin=0, ymax=29,
    xtick={2011,2012,2013,2014,2015,2016,2017},
    ytick={3,7,11,16,28},
    ymajorgrids=true,
    grid style=dashed,
]

\addplot[ color=blue, mark=square, ]
    coordinates { (2011,3)(2012,7)(2013,11)(2014,28)(2015,11)(2016,16)(2017,10) };
\end{axis}
\end{tikzpicture}
}
\caption{Number of studies over time.}
\label{fig:primary_studies_year}
\end{figure}
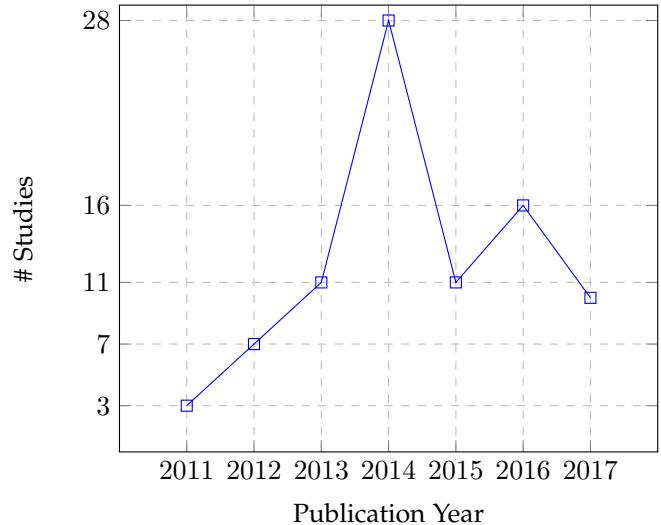

\figurename~\ref{fig:primary_studies_year} shows how the studies are distributed over time.
We can see that the interest in the topic of automated process discovery grew over time with a sharp increase between 2013 and 2014, and lately declining close to the average number of studies per year.

%We mapped these latter methods to their respective successors.
%Table~\ref{tab:superseded} summarize our findings. The only method without successor is \emph{AGNEs Miner}.
%As consequence, although we will refer to \emph{AGNEs Miner} in our benchmark,
%we will leave it out of from the SLR, since this method was already reviewed in previous work~\cite{DeWeerdt}.
%

% !TEX root = main.tex
\graphicspath{{plots/}}
\section{Classification of methods}
\label{sec:taxonomy}

% In this section, the results obtained from the SLR are presented.

%This indicates that automated process discovery has gained interest over the years. The fast grow of the number of research papers indicates the maturity of this field.
%
%\figurename~\ref{fig:primary_studies_year} shows how the primary studies are distributed over the years. It can be seen that, from year 2014 on, a growth in the number of papers has happened. This indicates that automated process discovery has gained interest over the years. The fast grow of the number of research papers indicates the maturity of this field. %The low number of studies included for years 2011 and 2012 can be explained by the fact that most of the approaches proposed in those years have already been updated or improved later on. The papers selected for 2016 are less than the ones selected for 2015 because we queried the data sources in October 2016 (as specified in Section~\ref{subsec:study selection}). This distribution also shows that process discovery is a hot theme in the field of process mining.
%
% In the remaining of this section we show the detailed results for our SLR answering each research question separately (from Section~\ref{subsec:rq1} to Section~\ref{subsec:rq6}).

%In this section, we show the detailed results for our SLR answering each research question separately (from Section~\ref{subsec:rq1} to Section~\ref{subsec:rq6}).

Driven by the research questions defined in Section~\ref{subsec:research_question}, we identified the following classification dimensions to survey the methods described in the primary studies:

%name of the discovery approach investigated, short list of authors and bibliography reference key, year of publication.

\begin{enumerate}
	\item Model type (procedural, declarative, hybrid) and model language (e.g., Petri nets, BPMN, Declare) --- RQ2
    \item Semantic captured in procedural models: parallelism (AND), exclusive choice (XOR), inclusive choice (OR), loop --- RQ3
	\item Type of implementation (standalone or plug-in, and tool accessibility) --- RQ4
	\item Type of evaluation data (real-life, synthetic or artificial log, where a synthetic log is one generated from a real-life model while an artificial log is one generated from an artificial model) and application domain (e.g., insurance, banking, healthcare) --- RQ5. %number, type, and size of logs tested, number of experiments performed.
\end{enumerate}

This information is summarized in Table~\ref{tab:SumTabel}.
Each entry in this table refers to the main study of the 35 groups found. Also, we cited all the studies that relate to the main one.
Collectively, the information reported by Table~\ref{tab:SumTabel} allows us to answer the first research question:
``what methods exist for automated process discovery?''

In the remainder of this section, we proceed with surveying each main study method along the above classification dimensions,
to answer the other research questions.

%provides an aggregated view of the most relevant information extracted from any of the selected primary studies. Such information was used %to further analyze any primary study
%for answering to the research questions, as detailed in Section \ref{sec:results}.

\begin{table*}[t!]
\small
\begin{adjustbox}{width=1\textwidth,center=\textwidth}
\pgfplotstabletypeset[
%column type=l,
every head row/.style={
before row={%
\toprule
\multicolumn{1}{c}{\textbf{Method}} &
\multicolumn{1}{c}{\textbf{Main study}} &
\multicolumn{1}{c}{\textbf{Year}} &
\multicolumn{1}{c}{\textbf{Related studies}} &
\multicolumn{1}{c}{\textbf{Model type}} &
\multicolumn{1}{c}{\textbf{Model language}} &
\multicolumn{4}{c}{\textbf{Semantic Constructs}} &
\multicolumn{2}{r}{\textbf{Implementation~~~~~~~}} &
\multicolumn{3}{c}{\textbf{Evaluation}} \\
},
after row=\midrule,
},
every last row/.style={
after row=\bottomrule},
columns/person/.style ={column name=},
columns/extra/.style ={column name=},
columns/authors/.style ={column name=},
columns/year/.style ={column name=},
columns/citations/.style ={column name=},
columns/relworks/.style ={column name=},
columns/modtype/.style ={column name=},
columns/langClass/.style ={column name=},
columns/para/.style ={column name=AND},
columns/exclusive/.style ={column name=XOR},
columns/or/.style ={column name=OR},
columns/loop/.style ={column name=Loop},
columns/impleF/.style ={column name=Framework},
columns/impleA/.style ={column name=Accessible},
columns/evReal/.style ={column name=Real-life},
columns/evSynth/.style ={column name=Synth.},
columns/evArt/.style ={column name=Art.},
%columns/forFree/.style ={column name=},
%columns/assessed/.style ={column name=},
col sep=&,row sep=\\,
string type,
]{
%person &
extra  & authors & year & relworks & modtype & langClass & para & exclusive & or & loop & impleF & impleA & evReal & evSynth & evArt\\
%%%%%%
%%%%%%
%%%%%%
HK & Huang and Kumar\cite{huang2012study} & 2012 & & Procedural & Petri nets & \checkmark & \checkmark & & \checkmark & Standalone & & \checkmark & \checkmark &  \\
Declare Miner & Maggi et al.\cite{DBLP:conf/caise/MaggiBA12} & 2012 & \cite{maggi2011user,maggi2013knowledge,DBLP:conf/cidm/BernardiCM14,DBLP:conf/bir/Maggi14,kala2016apriori,Maggi2017} & Declarative & Declare & & & & & ProM & \checkmark & \checkmark & \checkmark  &  \\
MINERful & Di Ciccio, Mecella\cite{di2013two} & 2013 & \cite{DBLP:conf/bis/CiccioM12,DBLP:journals/tmis/CiccioM15,di2014discovering,di2016efficient} & Declarative & Declare & & & & & ProM, Standalone & \checkmark & \checkmark  & \checkmark &  \\
Inductive Miner - Infrequent & Leemans et al.\cite{DBLP:conf/bpm/LeemansFA13} & 2013 & \cite{leemans2013discovering,leemans2014discovering,leemans2014exploring,leemans2015using,leemans2015scalable,leemansscalable,leemans2017modeling} & Procedural & Process trees & \checkmark & \checkmark & \checkmark & \checkmark & ProM & \checkmark & \checkmark &  &  \\
Data-aware Declare Miner & Maggi et al.\cite{maggi2013discovering} & 2013 & & Declarative & Declare & & & & & ProM & \checkmark & \checkmark & &  \\ %BPIC2011
Process Skeletonization & Abe, Kudo\cite{abe2014business} & 2014 & \cite{kudo2013business} & Procedural & Directly-follows graphs & & \checkmark &  & \checkmark & Standalone  & & \checkmark &  &   \\
Evolutionary Declare Miner & vanden Broucke et al.\cite{vanden2014declarative} & 2014 & & Declarative & Declare & & & & & Standalone & & & \checkmark & \\
Evolutionary Tree Miner & Buijs e al.\cite{buijs2014quality} & 2014 & \cite{van2011towards,buijs2012genetic,buijs2012role,buijs2013discovering,van2014genetic} & Procedural & Process trees & \checkmark & \checkmark & \checkmark &\checkmark  & ProM & \checkmark & \checkmark &  & \checkmark \\
Aim & Carmona, Cortadella\cite{carmona2014process} & 2014 &  & Procedural & Petri nets & \checkmark & \checkmark & & \checkmark & Standalone  & & \checkmark &  & \checkmark  \\
WoMan & Ferilli\cite{ferilli2014woman} & 2014 & \cite{ferilli2013logic,de2014learning,ferilli2014learning,ferilli2015logic,ferilli2016woman} & Declarative & WoMan & & & & & Standalone &  & \checkmark &  & \checkmark \\
Hybrid Miner & Maggi et al.\cite{maggi2014automated} & 2014 & & Hybrid & Declare + Petri nets &  &  &  &  & ProM & \checkmark & \checkmark &  &  \\
Competition Miner & Redlich et al.\cite{redlich2014scalable} & 2014 & \cite{redlich2014constructs,redlich2014introducing,redlich2014dynamic} & Procedural & BPMN & \checkmark & \checkmark &  & \checkmark & Standalone & & \checkmark & & \checkmark\\
Direted Acyclic Graphs & Vasilecas et al.\cite{vasilecas2014directed} & 2014 & & Procedural & Directed acyclic graphs & & \checkmark &  & & Standalone & & \checkmark & \checkmark &  \\
Fusion Miner & De Smedt et al.\cite{de2015fusion} & 2015 & & Hybrid & Declare + Petri nets &  &  &  &  & ProM & \checkmark & \checkmark & \checkmark & \checkmark \\
CNMining & Greco et al.\cite{greco2015process} & 2015 & \cite{greco2012process} & Procedural & Causal nets & \checkmark & \checkmark &  & \checkmark & ProM & \checkmark & \checkmark & \checkmark & \\
alpha\$ & Guo et al.\cite{guo2015mining} & 2015 & & Procedural & Petri nets & \checkmark & \checkmark &  & \checkmark & ProM & \checkmark & \checkmark &  & \checkmark \\
Maximal Pattern Mining & Liesaputra et al.\cite{liesaputra2015efficient} & 2015 & & Procedural & Causal nets & \checkmark & \checkmark &  & \checkmark & ProM & & \checkmark & & \checkmark \\
DGEM & Molka et al.\cite{molka2015diversity} & 2015 & & Procedural & BPMN & \checkmark & \checkmark &  & \checkmark & Standalone & & \checkmark & & \checkmark \\
ProDiGen & Vazquez et al.\cite{vazquez2015prodigen} & 2015 & \cite{vazquez2014genetic} & Procedural & Causal nets & \checkmark & \checkmark &  & \checkmark & ProM &  &  & \checkmark & \checkmark \\
Non-Atomic Declare Miner & Bernardi et al.\cite{bernardi2016activity} & 2016 & \cite{DBLP:conf/ruleml/BernardiCFM14} & Declarative & Declare & & & & & ProM & \checkmark & \checkmark & \checkmark & \\
RegPFA & Breuker et al.\cite{breuker2016comprehensible} & 2016 & \cite{breuker2014designing} & Procedural & Petri nets & \checkmark & \checkmark &  & \checkmark & Standalone & \checkmark & \checkmark & \checkmark & \\
BPMN Miner & Conforti et al.\cite{conforti2016bpmn} & 2016 & \cite{conforti2014beyond} & Procedural & BPMN & \checkmark & \checkmark & \checkmark & \checkmark & Apromore, Standalone & \checkmark & \checkmark & & \checkmark \\
CSMMiner & van Eck et al.\cite{van2016discovering} & 2016 & \cite{DBLP:conf/wecwis/EckSA17} & Procedural & State machines & \checkmark & \checkmark &  & \checkmark & ProM & \checkmark & \checkmark & & \\
TAU miner & Li et al.\cite{li2016process} & 2016 & & Procedural & Petri nets & \checkmark & \checkmark &  & \checkmark & ProM & & \checkmark & & \checkmark  \\
PGminer & Mokhov et al.\cite{mokhov2016mining} & 2016 & & Procedural & Partial order graphs & \checkmark & \checkmark &  & & Standalone, Workcraft & \checkmark & \checkmark & & \checkmark \\
SQLMiner & Sch\"{o}nig et al.\cite{schonig2016efficient} & 2016 & \cite{schonig2016discovery} & Declarative & Declare & & &  & & Standalone & \checkmark & \checkmark &  &  \\
ProM-D & Song et al.\cite{song2016process} & 2016 & & Procedural & Petri nets & \checkmark & \checkmark & & \checkmark & Standalone & & \checkmark & \checkmark & \\
CoMiner & Tapia-Flores et al.\cite{tapia2016discovering} & 2016 & & Procedural & Petri nets & \checkmark & \checkmark &  & \checkmark & ProM & & & & \checkmark \\
Proximity Miner & Yahya et al.\cite{yahya2016domain} & 2016 & \cite{yahya2013process} & Procedural & %Heuristic nets
Causal nets & \checkmark & \checkmark &  & \checkmark & ProM & \checkmark & \checkmark &  &  \\
Heuristics Miner & Augusto et al.\cite{augusto2017automated} & 2017 & \cite{weijters2011flexible,de2014improving,augusto2016automated,mannhardt2017data} & Procedural & BPMN & \checkmark & \checkmark &  & \checkmark & Apromore, Standalone & \checkmark & \checkmark & \checkmark & \checkmark \\
Split miner & Augusto et al.\cite{DBLP:conf/icdm/AugustoCDR17} & 2017 & & Procedural & BPMN & \checkmark & \checkmark & & \checkmark & Apromore, Standalone & \checkmark & \checkmark & & \\
Fodina & vanden Broucke et al.\cite{vanden2017fodina} & 2017 & \cite{de2014bidimensional} & Procedural & BPMN & \checkmark & \checkmark &  & \checkmark & ProM & \checkmark & \checkmark & \checkmark & \\
Stage miner & Nguyen et al.\cite{nguyen2017mining} & 2017 & & Procedural & Causal nets & \checkmark & \checkmark &  & \checkmark & Apromore, Standalone & \checkmark & \checkmark & & \\
Decomposed Process Miner & Verbeek, van der Aalst\cite{verbeek2017divide} & 2017 & \cite{verbeek2012experimental,van2013decomposing,hompes2014finding,verbeek2014decomposed,van2014process} & Procedural & Petri nets & \checkmark & \checkmark & \checkmark & \checkmark & ProM & \checkmark & \checkmark &  & \checkmark \\
HybridILPMiner & van Zelst et al.\cite{van2017discovering} & 2017 & \cite{van2015avoiding,van2015ilp} & Procedural & Petri nets & \checkmark & \checkmark &  & \checkmark & ProM & \checkmark &  &  & \checkmark \\
%%%%%%
}
\end{adjustbox}
\\ \\
\caption{Overview of the 35 primary studies resulting from the search (ordered by year and author).}
\label{tab:SumTabel}
\end{table*}

\subsection{Model type and language (RQ2)}
\label{subsec:rq2}

The majority of methods (26 out of 35) produce procedural models. Six approaches \cite{DBLP:conf/caise/MaggiBA12,di2013two,maggi2013discovering,vanden2014declarative,bernardi2016activity,schonig2016efficient} discover declarative models in the form of Declare constraints, whereas~\cite{ferilli2014woman} produces declarative models using the WoMan formalism. The methods in \cite{maggi2014automated,de2015fusion} are able to discover hybrid models as a combination of Petri nets and Declare constraints.

Regarding the modeling languages of the discovered process model, we notice that Petri nets is the predominant one. However, more recently, we have seen the appearance of methods that produce models in BPMN, a language that is more practically-oriented and less technical than Petri nets. This denotes a shift in the target audience of these methods, from data scientists to practitioners, such as business analysts and decision managers. Other technical languages employed, besides Petri nets, include Causal nets, State machines and simple Directed Acyclic Graphs, while Declare is the most commonly-used language when producing declarative models.

\medskip\noindent\textbf{Petri nets.} In~\cite{huang2012study}, the authors describe an algorithm to extract block-structured Petri nets from event logs. The algorithm works by first building an adjacency matrix between all pairs of tasks and then analyzing the information in it to extract block-structured models consisting of basic sequence, choice, parallel, loop, optional and self-loop structures as building blocks. The method has been implemented in a standalone tool called HK.

The method presented in~\cite{guo2015mining} is based on the $\alpha\$$ algorithm, which can discover invisible tasks involved in non-free-choice constructs. The algorithm is an extension of the well-known $\alpha$ algorithm, one of the very first algorithms for automated process discovery, originally presented in~\cite{DBLP:journals/tkde/AalstWM04}.

In~\cite{verbeek2017divide}, the authors propose a generic divide-and-conquer framework for the discovery of process models from large event logs.
The method allows to part the event log in smaller logs and discover a model from each of them. The output is then assembled from all the models discovered from the sublogs. The method aims to produce high quality models reducing overall the complexity. The preliminary studies \cite{verbeek2012experimental,van2013decomposing,hompes2014finding,verbeek2014decomposed,van2014process} widely illustrate the idea of splitting a large event log into a collection of smaller logs to improve the performance of a discovery algorithm.

van Zelst et al.~\cite{van2017discovering,van2015avoiding,van2015ilp} propose an improvement of the ILP miner implemented in~\cite{van2009process}, their method is based on hybrid variable-based regions. Through hybrid variable-based regions, it is possible to vary the number of variables used within the ILP problems being solved. Using a different number of variables has an impact on the average computation time for solving the ILP problem.

In~\cite{breuker2016comprehensible,breuker2014designing}, the authors propose an approach that allows the discovery of Petri nets using the theory of grammatical inference. The method has been implemented as a standalone application called RegPFA.

%Differently from traditional process discovery techniques that require stringent completeness notions of event logs to identify concurrency effectively,
The approach proposed in~\cite{song2016process} is based on the observation that activities with no dependencies in an event log can be executed in parallel. In this way, this method can discover process models with concurrency even if the logs fail to meet the completeness criteria. The method has been implemented in a tool called ProM-D.

In \cite{carmona2014process}, the authors propose the use of numerical abstract domains for discovering Petri nets from large logs while guaranteeing formal properties of the discovered models. The technique guarantees the discovery of Petri nets that can reproduce every trace in the log and that are minimal in describing the log behavior.

The approach introduced in \cite{tapia2016discovering} addresses the problem of discovering sound Workflow nets from incomplete logs. The method is based on the concept of invariant occurrence between activities, which is used to identify sets of activities (named conjoint occurrence classes) that can be used to infer the behaviors not exhibited in the log.

In \cite{li2016process}, the authors leverage data carried by tokens in the execution of a business process to track the state changes in the so-called token logs. This information is used to improve the performance of standard discovery algorithms.

\medskip\noindent\textbf{Process trees.} The Inductive Miner~\cite{DBLP:conf/bpm/LeemansFA13} and the Evolutionary Tree Miner~\cite{buijs2014quality} are both based on the extraction of process trees from an event log. Concerning the former, many different variants have been proposed during the last years, but its first appearance is in~\cite{leemans2013discovering}. Successively, since that method was unable to deal with infrequent behavior, an improvement was proposed in~\cite{DBLP:conf/bpm/LeemansFA13}, which efficiently drops infrequent behavior from logs, still ensuring that the discovered model is behaviorally correct (sound) and highly fitting.
Another variant of the Inductive Miner is presented in~\cite{leemans2014discovering}. This variant can minimize the impact of incompleteness of the input logs. In \cite{leemans2015using}, the authors discuss ways of systematically treating lifecycle information in the discovery task. They introduce a process discovery technique that is able to handle lifecycle data to distinguish between concurrency and interleaving. The method proposed in~\cite{leemans2014exploring} provides a graphical support for navigating the discovered model and the one described in \cite{leemans2017modeling} can deal with cancelation or error-handling behaviors (i.e., with logs containing traces that do not complete normally). %, while the one introduced in~\cite{leemans2015using}
Finally, the variant presented in~\cite{leemans2015scalable} and \cite{leemansscalable} combines scalability with quality guarantees. It can be used to mine large event logs and produces sound models. 

In~\cite{buijs2014quality}, Buijs et al.\ introduce the Evolutionary Tree Miner. This method is
based on a genetic algorithm that allows the user to drive the discovery process based on preferences with respect to the four quality dimensions of the discovered model: fitness, precision, generalization and complexity. The importance of these four dimensions and how to address their balance in process discovery is widely discussed in the related studies\cite{van2011towards,buijs2012genetic,buijs2012role,buijs2013discovering,van2014genetic}.
\medskip\noindent\textbf{Causal nets.}
Greco et al. propose a discovery method that returns causal nets \cite{greco2015process,greco2012process}. A causal net is a net where only the causal relation between activities in a log is represented. The proposed method encodes causal relations gathered from an event log and if available, background knowledge in terms of precedence constraints over the topology of the resulting process models. A discovery algorithm is formulated in terms of reasoning problems over precedence constraints.

In~\cite{liesaputra2015efficient}, the authors propose a method for automated process discovery using Maximal Pattern Mining where they discover recurrent sequences of events in the traces of the log. Starting from these patterns they build process models in the form of causal nets.

ProDiGen, a standalone miner by Vazquez et al.\cite{vazquez2015prodigen,vazquez2014genetic}, allows users to discover causal nets from event logs using a genetic algorithm. The algorithm is based on a fitness function that takes into account completeness, precision and complexity and specific crossover and mutation operators.
%Genetic can be used to derive a Heuristics net from the log. ProDiGen, a standalone miner by Vazguez et al.\cite{vazquez2015prodigen} represents this idea. It is basically a modified Genetic Miner[cite] and meant for control-flow discovery. New operators for crossover (selection based on the errors of the mined model) and mutation (relies on the log's causal dependencies) are specified, and also a new hierarchical fitness function that considers also completeness, precision (log and model based) and simplicity (model based), is used. Discovery is done in three-stages, at first log is pre-processed for removing the noise. Then, a model is mined, by using genetic approach and finally, the model is post-processed by removing infrequent and redundant arcs. Initial population is created by using the results of Heuristics Miner[cite]. Population individuals are evaluated with the fitness function and if the best match is found or the maximum number of reinitializations is reached, the process is stopped. The approach is available as a web-based front-end.

Another method that produces causal nets is the Proximity Miner, presented in~\cite{yahya2016domain,yahya2013process}. This method extracts behavioral relations between the events of the log which are then enhanced using inputs from domain experts.

Finally, in \cite{nguyen2017mining}, the authors propose a method to discover causal nets that optimizes the scalability and interpretability of the outputs. The process under analysis is decomposed into a set of stages, such that each stage can be mined separately. The proposed technique discovers a stage decomposition that maximizes modularity.

\medskip\noindent\textbf{State machines.}
%The authors of Process Spaceship \cite{motahari2011event} start from the observation that information about process executions is often scattered across several systems and data sources. Accordingly, they investigate different ways in which process-related events could be correlated in service interaction logs and propose a mechanism to discover event correlations (semi-)automatically from them. The data collected through event correlations is mined to discover process models in the form of state machines. This method has been implemented as an Eclipse plug-in.
%Finite state machines are used here to describe a mined log, thus it is a procedural technique. This approach uses event correlation for mining a log. This tool can be used in automated or in semi-automated mode. In the automated mode, the tool discovers a model by on a click. In the semi-automated mode, the user supervises the discovery and can select candidate attributes and conditions. The limitations for this approach is not catching concurrency, not tackling imperfect logs and not handling long traces. \\ \\
%There is also a third approach, that uses state machines, but it is categorizied as Others and therefore it will be described in the Section \ref{subsubsec:sm2}.
The CSM Miner, discussed in~\cite{van2016discovering,DBLP:conf/wecwis/EckSA17}, discovers state machines from event logs. Instead of focusing on the events or activities that are executed in the context of a particular process, this method concentrates on the
states of the different process perspectives and discover how they are related with each other. These relations are expressed in terms of Composite State Machines. The CSM Miner provides an interactive visualization of these multi-perspective state-based models.
%With van Eck et al.\cite{van2016discovering} approach, Composite State Machines(CSMs) can be discovered. The tool has been implemented as ProM plug-in and is called CSM Miner. CSMs can express sequence, parallel, exclusive choice and loops. The focus is on discovering states. Time for the log is not used for discovery in this approach, it is only used for adding statics at visualization stage. So a transition system is created from log, from which the CSMs are extracted. CSMs are further simplified, by removing redundant arcs, abstracting states and aggregating two given states into one. Support, confidence and lift are used to quantify the behavioural relations and can be used for further simplifying the model. Thus models can be interactively explored. The input is still an event log.

% \subsection{BPMN}
% \label{subsubsec:bpmn}
%In~\cite{conforti2014beyond,conforti2016bpmn,augusto2016automated,redlich2014dynamic,molka2015diversity,van2015process} approaches for discovering BPMN models from event logs are presented.

\medskip\noindent\textbf{BPMN models.} In~\cite{conforti2014beyond}, Conforti et al.\ present the BPMN Miner, a method for the automated discovery of BPMN models containing sub-processes, activity markers such as multi-instance and loops, and interrupting and non-interrupting boundary events (to model exception handling).
%The technique analyzes dependencies between data attributes attached to events in order to identify sub-processes and to extract a set of corresponding sub-logs. Parent process and sub-process models are then discovered using existing techniques for flat process model discovery. The models can be, for example Petri nets that can be easily converted into BPMN models. In a final stage, the resulting models and logs are heuristically analyzed in order to identify boundary events and markers.
The method has been subsequently improved in~\cite{conforti2016bpmn} to make it robust to noise in event logs. 
%In order to handle noisy logs, the improved technique employs approximate dependency discovery techniques, followed by a series of filters designed to remove noise at the lowest possible level of granularity. Approximate dependency discover techniques can discover a sub-process even if not all instances of this sub-process refer correctly to their corresponding parent process. This approach has been implemented in an application called BPMN Miner, which is available as a standalone application and, also, as a ProM plug-in and an Apromore plug-in.
%in 2015 to detect and filter noise. The BPMN Miner is available as ProM or Apromore plug-in and also as standalone. The noise handling is done by using approximate dependency discovery techniques and a set of filters. To produce a model, the set of events of a event type is seen as relational table, likelihood of events that share same primary key will be long to the same process and that process-subprocess relations are indicated by foreign keys between event types are also used. At first a flat-model is mined with Heuristic Miner, ILP, InductiveMiner, Fodina Heuristic Miner or $\alpha$-algorithm by user choice. Then this flat-model is re factored by using a set of heuristics to be hierarchical. BPMN Miner works best with logs that contain records of a single business process. \\ \\

Another method to discover BPMN models has been presented in \cite{molka2015diversity}. In this approach, a hierarchical view on process models is formally specified and an evolution strategy is applied on it. The evolution strategy, which is guided by the diversity of the process model population, efficiently finds the process models that best represent a given event log.

A further method to discover BPMN models is the Dynamic Constructs Competition Miner~\cite{redlich2014scalable,redlich2014introducing,redlich2014dynamic}. This method extends the Constructs Competition Miner presented in~\cite{redlich2014constructs}. The method is based on a divide-and-conquer algorithm which discovers block-structured process models from logs. 

In \cite{DBLP:conf/icdm/AugustoCDR17}, the authors propose a discovery method that produces simple process models with low branching complexity and consistently high and balanced fitness, precision and generalization. The approach combines a technique to filter the directly-follows graph induced by an event log, with an approach to identify combinations of split gateways that accurately capture the concurrency, conflict and causal relations between neighbors in the directly-follows graph.

Fodina \cite{vanden2017fodina,de2014bidimensional} is a discovery method based on the Heuristics Miner~\cite{weijters2006process}.
However, differently from the Heuristics Miner, Fodina is more robust to noisy data, is able to discover duplicate activities,
and allows for flexible configuration options to drive the discovery according to end user inputs.

In~\cite{weijters2011flexible}, the authors present the Flexible Heuristics Miner. This method can discover process models containing non-trivial constructs but with a low degree of block structuredness. At the same time, the method can cope well with noise in event logs. The discovered models are a specific type of Heuristics nets where the semantics of splits and joins is represented using split/join frequency tables. This results in easy to understand process models even in the presence of non-trivial constructs and log noise. The discovery algorithm is based on that of the original Heuristics Miner method~\cite{weijters2006process}. In~\cite{de2014improving}, the method presented in~\cite{weijters2011flexible} has been improved as anomalies were found concerning the validity and completeness of the resulting process model. The improvements have been implemented in the Updated Heuristics Miner. A data-aware version of the Heuristics Miner that takes into consideration data attached to events in a log has been presented in \cite{mannhardt2017data}. Finally, in \cite{augusto2017automated,augusto2016automated}, the authors propose an improvement of the Heuristics Miner algorithm to separate the objective of producing accurate models and that of ensuring their structuredness and soundness. Instead of directly discovering a structured process model, the approach first discovers accurate, possibly unstructured (and unsound) process models, and then transforms the resulting process model into a structured (and sound) one.

\medskip\noindent\textbf{Declarative models.} In \cite{maggi2011user}, the authors present the first basic approach for mining declarative process models expressed using Declare constraints~\cite{DBLP:conf/edoc/PesicSA07,DBLP:conf/bpm/WestergaardM11}. This approach was improved in \cite{DBLP:conf/caise/MaggiBA12} using a two-phase approach. The first phase is based on an apriori algorithm used to identify frequent sets of correlated activities. A list of candidate constraints is built on the basis of the correlated activity sets. In the second phase, the constraints are checked by replaying the log on specific automata, each accepting only those traces that are compliant to one constraint. Those constraints satisfied by a percentage of traces higher than a user-defined threshold, are discovered. Other variants of the same approach are presented in \cite{maggi2013knowledge,DBLP:conf/cidm/BernardiCM14,DBLP:conf/bir/Maggi14,kala2016apriori,Maggi2017}. The technique presented in \cite{maggi2013knowledge} leverages apriori knowledge to guide the discovery task. In \cite{DBLP:conf/cidm/BernardiCM14}, the approach is adapted to be used in cross-organizational environments in which different organizations execute the same process in different variants. In \cite{DBLP:conf/bir/Maggi14}, the author extends the approach to discover metric temporal constraints, i.e., constraints taking into account the time distance between events. Finally, in \cite{kala2016apriori,Maggi2017}, the authors propose mechanisms to reduce the execution times of the original approach presented in \cite{DBLP:conf/caise/MaggiBA12}.

MINERful~\cite{di2013two,DBLP:conf/bis/CiccioM12,DBLP:journals/tmis/CiccioM15} discovers Declare constraints using a two-phase approach. The first phase computes statistical data describing the occurrences of activities and their interplay in the log. The second one checks the validity of Declare constraints by querying such a statistic data structure (knowledge base). In \cite{di2014discovering,di2016efficient}, the approach is extended to discover target-branched Declare constraints, i.e., constraints in which the target parameter is the disjunction of two or more activities.

The approach presented in \cite{maggi2013discovering} is the first approach for the discovery of Declare constraints with an extended semantics that take into consideration data conditions. The data-aware semantics of Declare presented in this paper is based on first-order temporal logic. The method presented in~\cite{bernardi2016activity,DBLP:conf/ruleml/BernardiCFM14} is based on the use of discriminative rule mining to determine how the characteristics of the activity lifecycles in a business process influence the validity of a Declare constraint in that process.

Other approaches for the discovery of Declare constraints have been presented in \cite{vanden2014declarative,schonig2016efficient}. In \cite{vanden2014declarative}, the authors present the Evolutionary Declare Miner that implements the discovery task using a genetic algorithm.
%In~\cite{di2016efficient}, the authors present a method to discover a class of Declare constraints called target-branched Declare. A Declare constraint is target-branched when one of its parameters (the target) is the disjunction of two or more activities to express that one activity out of a set of activities can occur. The method has been implemented as a standalone application.
The SQLMiner, presented in \cite{schonig2016efficient}, is based on a mining approach that directly works on relational event data by querying a log with standard SQL. By leveraging database performance technology, the mining procedure is extremely fast. Queries can be customized and cover process perspectives beyond control flow\cite{schonig2016discovery}.

The WoMan framework, proposed by Ferilli in \cite{ferilli2014woman} and further extended in the related studies \cite{ferilli2013logic,de2014learning,ferilli2014learning,ferilli2015logic,ferilli2016woman}, includes a method to learn and refine process models from event logs, by discovering first-order logic constraints. It guarantees incrementality in learning and adapting the models, the ability to express triggers and conditions on the process tasks and efficiency.

\medskip\noindent\textbf{Further approaches.}  In~\cite{abe2014business,kudo2013business}, the authors introduce a monitoring framework for automated process discovery. A monitoring context is used to extract traces from relational event data and attach different types of metrics to them. Based on these metrics, traces with certain characteristics can be selected and used for the discovery of process models expressed as directly-follows graphs.

Vasilecas et al.\ ~\cite{vasilecas2014directed} present a method for the extraction of directed acyclic graphs from event logs. Starting from these graphs, they generate Bayesian belief networks, one of the most common probabilistic models, and use these networks to efficiently analyze business processes.

In~\cite{mokhov2016mining}, the authors show how conditional partial order graphs, a compact representation of families of partial orders, can be used for addressing the problem of compact and easy-to-comprehend representation of event logs with data. They present algorithms for extracting both the control flow as well as relevant data parameters from a given event log and show how conditional partial order graphs can be used to visualize the obtained results. The method has been implemented as a Workcraft plug-in and as a standalone application called PGminer.

The Hybrid Miner, presented in \cite{maggi2014automated}, puts forward the idea of discovering a hybrid model from an event log based on the semantics defined in~\cite{Slaats2016}. According to such semantics, a hybrid process model is a hierarchical model, where each node represents a sub-process, which may be specified in a declarative or procedural way. Petri nets are used for representing procedural sub-processes and Declare for representing declarative sub-processes.
%Mokhov et al.\cite{mokhov2016mining} created an approach Workcraft plug-in for extracting Conditional Partial Order Graphs(CPOG). It is also implemented as command line standalone tool named PGMINER. Sequence, parallel and exclusive choice constructs can be extracted from the logs. Control- and data-flows can be extracted. CPOGs can be mined by treating each trace as a totally ordered sequence of events or by exploiting the concurrency between the events. With both, all the traces are covered from the log. Event attributes are used for adding data labels to the conditions. The quadratic explosion of the representation is avoided by using the algebra of Parametrised Graphs. The input is an event log. If log is imported directly, each trace is treated as a total order of events. If imported indirectly via PGMINER, the log undergoes the concurrency extraction. Second option allows handling of bigger logs.

\cite{de2015fusion} proposes an approach for the discovery of hybrid models based on the semantics proposed in \cite{westergaard2013mixing}. Differently from the semantics introduced in~\cite{Slaats2016}, where a hybrid process model is hierarchical, the semantics defined in \cite{westergaard2013mixing} is devoted to obtain a fully mixed language, where procedural and declarative constructs can be connected with each other.

\subsection{Procedural language constructs (RQ3)}
\label{subsec:rq3}

All the 26 methods that discover a procedural model can detect the basic control-flow structure of sequence. Out of these methods, only four can also discover inclusive choices, but none in the context of non-block-structured models. In fact, \cite{DBLP:conf/bpm/LeemansFA13,buijs2014quality} are able to
directly identify block-structured inclusive choices (using process trees), while~\cite{verbeek2017divide,conforti2016bpmn} can detect this construct only when used on top of the methods in~\cite{buijs2014quality} or~\cite{DBLP:conf/bpm/LeemansFA13} (i.e., indirectly).

The remaining 22 methods can discover constructs for parallelism, exclusive choice and loops, with the exception of \cite{abe2014business}, which can detect exclusive choice and loops but not parallelism, \cite{mokhov2016mining}, which can detect parallelism and exclusive choice but not loops, and \cite{vasilecas2014directed}, which can discover exclusive choices only.
 %This is mostly due to the nature of their outputs.
%Indeed, the languages supported by these methods, i.e., directly-follows graphs, heuristics nets, directed acyclic graphs and casual nets do not natively support parallelism. In other methods these languages have been extended to cater for parallelism (cf.~\cite{greco2015process,vazquez2015prodigen,yahya2016domain}).

%Even if not explicitly mentioned in Table \ref{tab:SumTabel}, it is worth notice that any of the approaches investigated allow the detection of simple sequences within the discovered process models.

% About the declarative discovery approaches, we cannot determine the same constructs as in the case of procedural models. In this case we considered the discoverable declarative patterns.
% %
% In particular, the approaches~\cite{DBLP:conf/caise/MaggiBA12,di2013two,bernardi2016activity,di2016efficient,schonig2016efficient} are able to identify the whole set of Declare constraints, while~\cite{ferilli2014woman} detects constructs belonging to the WoMan formalism.

%, which all of these latter approaches are able to discover completely (i.e., the whole set of Declare patterns). 

\subsection{Implementation (RQ4)}
\label{subsec:rq4}

Over 50\% of the methods (19 out of 35) provide an implementation as a plug-in for the ProM platform.~\footnote{\url{http://promtools.org}} The reason behind the popularity of ProM can be explained by its open-source and portable framework, which allows researchers to easily develop and test new discovery algorithms. Also, ProM is the first software tool for process mining. One of the methods which has a ProM implementation \cite{di2013two} is also available as standalone tool. The works~\cite{conforti2016bpmn,DBLP:conf/icdm/AugustoCDR17,nguyen2017mining,augusto2017automated} provide both a standalone implementation and a further implementation as a plug-in for Apromore.\footnote{\url{http://apromore.org}} Apromore is an online process analytics platform, also open source, and has a growing consensus among academics as a process mining tool oriented towards end users. Finally, one method \cite{mokhov2016mining} has been implemented as a plug-in for Workcraft,~\footnote{\url{http://workcraft.org}} a platform for designing concurrent systems.

Notice that 22 tools out of 35 are made publicly available to the community. These exclude 4 ProM plug-ins and 9 standalone tools. 

\subsection{Evaluation data and domains (RQ5)}
\label{subsec:rq5}

The surveyed methods have been evaluated using three types of event logs: \emph{(i)} real-life logs, i.e., logs of real-life process execution data; \emph{(ii)} synthetic logs, generated by replaying real-life process models; and \emph{(iii)} artificial logs, generated by replaying artificial models.

We found that the majority of methods (31 out of 35) were tested using real-life logs. Among them, 11 approaches (cf. \cite{huang2012study,DBLP:conf/caise/MaggiBA12,di2013two,vasilecas2014directed,greco2015process,de2015fusion,song2016process,breuker2016comprehensible,bernardi2016activity,vanden2017fodina,augusto2017automated}) were further tested against synthetic logs, while 13 approaches (cf. \cite{carmona2014process,redlich2014scalable,buijs2014quality,ferilli2014woman,guo2015mining,molka2015diversity,liesaputra2015efficient,de2015fusion,conforti2016bpmn,mokhov2016mining,li2016process,verbeek2017divide,augusto2017automated}) against artificial logs. Finally, one method was tested both on synthetic and artificial logs only (cf. \cite{vazquez2015prodigen}), while~\cite{tapia2016discovering,van2017discovering} were tested on artificial logs and \cite{vanden2014declarative} on synthetic logs only.
Among the methods that employ real-life logs, we observed a growing trend in employing publicly-available logs, as opposed to private logs which hamper the replicability of the results due to not being accessible.

%\todo[inline]{Should we say something here about the size and the noise of the logs tested? Consider that such information is already tackled in the quality assessment and in the benchmark section.}

Concerning the application domains of the real-life logs, we noticed that several methods used a selection of the logs made available by the Business Process Intelligence Challenge (BPIC), which is held annually as part of the BPM Conference series. These logs are publicly available at the \emph{4TU Centre for Research Data},~\footnote{\url{https://data.4tu.nl/repository/collection:event_logs_real}} and cover domains such as healthcare (used by~\cite{maggi2013discovering,di2013two,DBLP:conf/bpm/LeemansFA13,liesaputra2015efficient,schonig2016efficient,DBLP:conf/icdm/AugustoCDR17,augusto2017automated}), banking (used by~\cite{di2013two,DBLP:conf/bpm/LeemansFA13,maggi2014automated,vasilecas2014directed,molka2015diversity,breuker2016comprehensible,schonig2016efficient,van2016discovering,DBLP:conf/icdm/AugustoCDR17,augusto2017automated,verbeek2017divide}), IT support management in automotive (cf. \cite{bernardi2016activity,breuker2016comprehensible,DBLP:conf/icdm/AugustoCDR17,augusto2017automated}), and public administration (cf. \cite{DBLP:conf/caise/MaggiBA12,DBLP:conf/bpm/LeemansFA13,carmona2014process,buijs2014quality,verbeek2017divide}). A public log pertaining to a process for managing road traffic fines (also available at the \emph{4TU Centre for Research Data}) was used in \cite{DBLP:conf/icdm/AugustoCDR17,augusto2017automated}. In \cite{mokhov2016mining}, the authors use logs from various domains available at \url{http://www.processmining.be/actitrac/}.

%was tested against a log extracted by an on-line game service, while \cite{huang2012study} against a public log describing patent application procedures in United States.

Besides these publicly-available logs, a range of private logs were also used, originating from different domains such as logistics (cf. \cite{song2016process,yahya2016domain}), traffic congestion dynamics~\cite{greco2015process}, employers habits (cf. \cite{ferilli2014woman,van2016discovering}), automotive~\cite{guo2015mining}, healthcare~\cite{molka2015diversity,DBLP:conf/icdm/AugustoCDR17,augusto2017automated}, and project management and insurance (cf. \cite{conforti2016bpmn,abe2014business}).

\section{Benchmark}
\label{sec:benchmark}

Using a selection of the methods surveyed in this paper, we conducted an extensive benchmark to identify relative advantages and tradeoffs. In this section, we justify the criteria of the methods selection, describe the datasets, the evaluation setup and metrics, and present the results of the benchmark. These results, consolidated with the findings from the systematic literature review, are then discussed in Section~\ref{sec:discussion}.

\subsection{Methods selection}

Assessing all the methods that resulted from the search would not be possible due to the heterogeneous nature of the inputs required and the outputs produced. Hence, we decided to focus on the largest subset of comparable methods. The methods considered were the ones satisfying the following criteria:
%This latter is identified by the following criteria:
%Lastly, we decided to run a benchmark on a subset of the discovery approaches analyzed for this literature review.
%Precisely, we filled the subset applying the following criteria:
\begin{itemize}
\item[i] an implementation of the method is publicly accessible; %and
\item[ii] the output of the method is a Petri net or a model seamlessly convertible into a Petri net (i.e., process trees and BPMN models). %and
% \item[iv] the corresponding paper counts more than 10 citations OR provides a \emph{strong evaluation} of the proposed method.
\end{itemize}

The second criterion is a requirement dictated by the metrics used to evaluate the accuracy of a discovered model (illustrated later in this section) that can only be computed on top of Petri nets.

The application of these criteria resulted in an initial selection of the following methods (corresponding to one third of the total studies):
%On the basis of these criteria, we initially selected the following methods (corresponding to one third of the total studies):
$\alpha\$$~\cite{guo2015mining},
Inductive Miner~\cite{leemans2013discovering},
Evolutionary Tree Miner~\cite{buijs2014quality},
Fodina~\cite{vanden2017fodina},
Structured Heuristic Miner 6.0~\cite{augusto2017automated},
Split Miner~\cite{DBLP:conf/icdm/AugustoCDR17},
Hybrid ILP Miner~\cite{van2015ilp},
RegPFA~\cite{breuker2016comprehensible},
Stage Miner~\cite{nguyen2017mining},
BPMN Miner~\cite{conforti2016bpmn},
Decomposed Process Mining~\cite{verbeek2014decomposed}.

A posteriori, we excluded the latter four due to the following reasons: Decomposed Process Mining, BPMN Miner, and Stage Miner were excluded as such approaches follow a divide-and-conquer approach which could be applied on top of any discovery method to improve its results; on the other hand, we excluded RegPFA because its output is a graphical representation of a Petri net (DOT), which could not be seamlessly serialized into the standard Petri net format.

We also considered including commercial process mining tools in the benchmark. Specifically, we investigated Disco\footnote{\url{http://fluxicon.com/disco}}, Celonis\footnote{\url{http://www.celonis.com}}, Minit\footnote{\url{http://minitlabs.com/}}, and myInvenio\footnote{\url{http://www.my-invenio.com}}. Disco and Minit are not able to produce business process models from event logs. Instead, they can only produce directly-follows graphs, which do not have an execution semantics. Indeed, when a given node (task) has several incoming arcs, a directly-follows graph does not tell us whether or not the task in question should wait for all its incoming tasks to complete, or just for one of them, or a subset of them. A similar remark applies to split points in the directly-follows graph. Given their lack of execution semantics, it is not possible to directly translate a directly-follows graph into a BPMN models or a Petri net.  Instead, one has to determine what is the intended behavior at each split and join point, which is precisely what several of the automated process discovery techniques based on directly-follows graph do (e.g., the Inductive Miner or Split Miner).

Celonis and myInvenio can produce BPMN process models but all they do is to insert OR (inclusive) gateways at the split and join points of the process map. The resulting BPMN process models cannot be translated into Petri nets using existing mappings from BPMN to Petri nets~\cite{Favre2015}. The Split Miner adopts a similar approach (it initially inserts OR-join gateways at join points in the directly-follows graph), but it then applies an algorithm to turn these OR-join gateways into combinations of XOR and AND gateways -- at the expense of potentially making the process model unsafe (but deadlock-free). It would have been possible to take the output of Celonis and myInvenio and apply an approach similar to the Split Miner to remove the OR-join gateways, but the resulting technique would be in essence the Split Miner itself, and the latter is already included in the benchmark.

%A posteriori, we excluded the latter four due to the following reasons. Decomposed Process Mining, BPMN Miner, and Stage Miner were excluded because such approaches follow a divide-and-conquer approach, and therefore they can be applied on top of any other discovery method to improve the results. Whilst, we excluded RegPFA because its output is a Petri net only available in graphical format (DOT).

%However, Decomposed Process Mining is a divide and conquer approach which can be applied on top of any other discovery approach, therefore we decided to not include it in our benchmark and consider it as a possible way to further improve the results of the other discovery approaches. Also, we had to exclude Maximal Pattern Mining and ProDiGen, since despite claimed the implementations in the respective papers, we were not able to find them and we got no reply from the authors.

In conclusion, the final selection of methods for the benchmark contained: $\alpha\$$, Inductive Miner (IM), Evolutionary Tree Miner (ETM), Fodina (FO), Structured Heuristic Miner 6.0 (S-HM\textsubscript{6}), Split Miner (SM), and Hybrid ILP Miner (HILP).%, latest version available.

\subsection{Setup and datasets}

To guarantee the reproducibility of our benchmark and to provide the community with a tool for comparing new methods with the ones evaluated in this paper, we developed a command-line Java application that performs measurements of accuracy and complexity metrics on the seven discovery methods selected above.
%This application takes as input a list of discovery approaches and a list of metrics to measure on the discovered process models. After the evaluation, the tool automatically exports all the results on an excel file.
The tool can be easily extended to incorporate new discovery methods and metrics.\footnote{Tool available at \url{https://github.com/raffaeleconforti/ResearchCode/releases/download/v0.1/Benchmark.v0.1.zip} and its source code is available at \url{https://github.com/raffaeleconforti/ResearchCode}.}

For our evaluation, we used two datasets. The first is the collection of real-life event logs publicly available at the 4TU Centre for Research Data as of March 2017.\footnote{\url{https://data.4tu.nl/repository/collection:event_logs_real}} Out of this collection, we considered the \emph{BPI Challenge} (BPIC) logs, the \emph{Road Traffic Fines Management Process} (RTFMP) log, and the \emph{SEPSIS Cases} log. These logs record executions of business processes from a variety of domains, e.g., healthcare, finance, government and IT service management. For our evaluation we held out those logs that do not explicitly capture business processes (i.e., the BPIC 2011 and 2016 logs), and those contained in other logs (e.g., the \emph{Environmental permit application process} log). Finally, in three logs (i.e., the BPIC14, BPIC15, and BPIC17 logs), we applied the filtering technique proposed in~\cite{ConfortiRH17} to remove infrequent behavior. This was necessary since all the models discovered by the considered methods exhibited very poor accuracy (F-score close to 0 or not computable) on these logs, making the comparison useless.

Table~\ref{tab:publogs} reports the characteristics of the twelve logs used. These logs are widely heterogeneous ranging from simple to very complex, with a log size ranging from 681 traces (for the BPIC15\textsubscript{2f} log) to 150.370 traces (for the RTFMP log). A Similar variety can be observed in the percentage of distinct traces, ranging from 0,2\% to 80,6\%, and the number of event classes (i.e., activities executed within the process), ranging from 7 to 82. Finally, the length of a trace also varies from very short, with traces containing only one event, to very long with traces containing 185 events.

The second dataset is composed of twelve proprietary logs sourced from several companies around the world. %NOT SURE WE CAN MENTION HITACHI AND SUNCORP WITHOUT ETHICAL CLEARANCE%among which we include Hitachi and Suncorp.
Table~\ref{tab:privlogs} reports the characteristics of these logs. Also in this case, the logs are quite heterogeneous, with the number of traces (and the percentage of distinct traces) ranging from 225 (of which 99,9\% distinct) to 787.657 (of which 0,01\% distinct). The number of recorded events varies between 4.434 and 2.099.835, whilst the number of event classes ranges from 8 to 310. %Similarly, also the shortest and the longest traces widely vary across the dataset.

\begin{table*}
	\centering
%	{\scriptsize{
    \begin{tabular}{c|cccc|ccc}
    
    \textbf{Log}
    & \textbf{Total}
    & \textbf{Distinct}
    & \textbf{Total}
    & \textbf{Distinct}
    & \multicolumn{3}{c}{\textbf{Trace Length}}
    \\\cline{6-8}
    \textbf{Name}
    & \textbf{Traces}
    & \textbf{Traces (\%)}
    & \textbf{Events}
    & \textbf{Events}
    & \textbf{min}
    & \textbf{avg}
    & \textbf{max} \\\hline
    
	BPIC12 & 13,087 & 33.4 & 262,200 & 36 & 3 & 20 & 175 \\\hline
	
	BPIC13\textsubscript{cp} & 1,487 & 12.3 & 6,660 & 7 & 1 & 4 & 35\\\hline
	
	BPIC13\textsubscript{inc} & 7,554 & 20.0 & 65,533 & 13 & 1 & 9 & 123\\\hline
	
	BPIC14\textsubscript{f} & 41,353 & 36.1 & 369,485 & 9 & 3 & 9 & 167 \\\hline
	
	BPIC15\textsubscript{1f} & 902 & 32.7 & 21,656 & 70 & 5 & 24 & 50 \\\hline
	
	BPIC15\textsubscript{2f} & 681 & 61.7 & 24,678 & 82 & 4 & 36 & 63 \\\hline
	
	BPIC15\textsubscript{3f} & 1,369 & 60.3 & 43,786 & 62 & 4 & 32 & 54 \\\hline
	
	BPIC15\textsubscript{4f} & 860 & 52.4 & 29,403 & 65 & 5 & 34 & 54 \\\hline
	
	BPIC15\textsubscript{5f} & 975 & 45.7 & 30,030 & 74 & 4 & 31 & 61 \\\hline
	
	BPIC17\textsubscript{f} & 21,861 & 40.1 & 714,198 & 41 & 11 & 33 & 113 \\\hline
	
	RTFMP & 150,370 & 0.2  & 561,470 & 11 & 2 & 4 & 20\\\hline
	
	SEPSIS & 1,050 & 80.6 & 15,214 & 16 & 3 & 14 & 185 \\\hline	

	\end{tabular}
%    }}
  	\caption{Descriptive statistics of the public event logs.}\label{tab:publogs}
\end{table*}

\begin{table*}
	\centering
%	{\scriptsize{
    \begin{tabular}{c|cccc|ccc}
    
    \textbf{Log}
    & \textbf{Total}
    & \textbf{Distinct}
    & \textbf{Total}
    & \textbf{Distinct}
    & \multicolumn{3}{c}{\textbf{Trace Length}}
    \\\cline{6-8}
    \textbf{Name}
    & \textbf{Traces}
    & \textbf{Traces (\%)}
    & \textbf{Events}
    & \textbf{Events}
    & \textbf{min}
    & \textbf{avg}
    & \textbf{max} \\\hline
    
	PRT1 & 12,720 & 8.1 & 75,353 & 9 & 2 & 5 & 64 \\\hline
	
	PRT2 & 1,182 & 97.5 & 46,282 & 9 & 12 & 39 & 276 \\\hline

	PRT3 & 1,600 & 19.9 & 13,720 & 15 & 6 & 8 & 9 \\\hline

	PRT4 & 20,000 & 29.7 & 166,282 & 11 & 6 & 8 & 36 \\\hline

	PRT5 & 739 & 0.01 & 4,434 & 6 & 6 & 6 & 6 \\\hline

	PRT6 & 744 & 22.4 & 6,011 & 9 & 7 & 8 & 21 \\\hline

	PRT7 & 2,000 & 6.4 & 16,353 & 13 & 8 & 8 & 11 \\\hline
	
	PRT8 & 225 & 99.9 & 9,086 & 55 & 2 & 40 & 350 \\\hline
	
	PRT9 & 787,657 & 0.01 & 1,808,706 & 8 & 1 & 2 & 58 \\\hline
	
	PRT10 & 43,514 & 0.01 & 78,864 & 19 & 1 & 1 & 15 \\\hline
	
	PRT11 & 174,842 & 3.0 & 2,099,835 & 310 & 2 & 12 & 804 \\\hline
	
	PRT12 & 37,345 & 7.5 & 163,224 & 20 & 1 & 4 & 27 \\\hline
	
	\end{tabular}
%    }}
  	\caption{Descriptive statistics of the proprietary event logs.}\label{tab:privlogs}
\end{table*}

We performed our benchmark on a 6-core Intel Xeon CPU E5-1650 v3 @ 3.50GHz with 128GB RAM running Java 8. We allocated a total of 16GB to the heap space, and we enforced a timeout of four hours for the discovery phase and one hour for measuring each of the quality metrics.

\subsection{Evaluation metrics}

For all selected discovery metrics we measured the following accuracy and complexity metrics: recall (a.k.a.\ fitness), precision, generalization, complexity, and soundness.

\emph{Fitness} measures the ability of a model to reproduce the behavior contained in a log. Under trace semantics, a fitness of 1 means that the model can reproduce every trace in the log. In this paper, we use the fitness measure proposed in~\cite{AdriansyahDA11}, which measures the degree to which every trace in the log can be aligned (with a small number of errors) with a trace produced by the model. In other words, this measures tells us how close on average a given trace in the log can be aligned with a trace that can be generated by the model.

\emph{Precision} measures the ability of a model to generate only the behavior found in the log. A score of 1 indicates that any trace produced by the model is contained in the log. In this paper, we use the precision measure defined in~\cite{AdriansyahMCDA12}, which is based on similar principles as the above fitness measure.
Recall and precision can be combined into a single measure known as F-score, which is the harmonic mean of the two measurements $\left( 2 \cdot \frac{\mathit{Fitness} \cdot \mathit{Precision}}{\mathit{Fitness} + \mathit{Precision}} \right)$.

\emph{Generalization} refers to the ability of an automated discovery algorithm to discover process models that generate traces that are not present in the log but that can be produced by the business process under observation. In other words, an automated process discovery algorithm has a high generalization on a given event log if it is able to discover a process model from the event log, which generates traces that: (i) are not in the event log, but (ii) can be produced by the business process that produced the event log. Note that unlike fitness and precision, generalization is a property of an algorithm on an event log, and not a property of the model produced by an algorithm when applied to a given event log.

In line with the above definition, we use k-fold cross-validation~\cite{Kohavi95} to measure event logs. This k-fold cross-validation approach to measure generalization has been advocated in several studies in the field of automated process discovery~\cite{Rozinat2007,ProcessMiningBook,Bolt2016,vanDongen2017}.
Concretely, we divide the log into $k$ parts, we discover a model from $k-1$ parts (i.e., we hold-out one part), and measure the fitness of the discovered model against the part held out. This is repeated for every possible part held out. Generalization is the mean of the fitness values obtained for each part held out. A generalization of one means that the discovered model produces traces in the observed process, even if those traces are not in the log from which the model was discovered, and that the discovered model is accurate and does not introduce extra behavior (i.e., does not over-generalize the behavior recorded in the log).

In the results reported below, we use a value of $k=3$ for performance reasons (as opposed to the classical value of $k=10$). The fitness calculation for most of the algorithm-log pairs is slow, and repeating it $10$ times, for every algorithm-log combination is costly.  To test if the results could be affected by this choice of $k$, we used $k=10$ for SM and IM on the BPIC12 log, and found that the value of the 10-fold generalization measure was within one percentage point of that of the 3-fold generalization measure.

%\emph{Generalization} is meant to measure the ability of a discovery approach to capture behavior not completely recorded in the log. To measure generalization we use 3-fold cross validation. Firstly, we divide the log into 3 parts, we then discover the model from 2 parts (i.e., we hold-out 1 part), and finally, we measure the fitness of the discovered model against the hold-out part, and precision of the discovered model against the complete log. This operation is repeated for every possible holdout part, and the average over the measurements constitute the \emph{3-fold fitness} and the \emph{3-fold precision}. Finally, we report the F-score of the 3-fold fitness and precision as unique measure for generalization.

\emph{Complexity} quantifies how difficult it is to understand a model. Several complexity metrics have been shown to be (inversely) related to understandability~\cite{Mendling08}, including \emph{Size} (number of nodes); \emph{Control-Flow Complexity (CFC)} (the amount of branching caused by gateways in the model) and \emph{Structuredness} (the percentage of nodes located directly inside a block-structured single-entry single-exit fragment).

Lastly, \emph{soundness} assesses the behavioral quality of a process model by reporting whether the model violates one of the three soundness criteria~\cite{aalst1997}: i) option to complete, ii) proper completion, and iii) no dead transitions.

\subsection{Benchmark results}

% !TEX root = main.tex
\begin{table*}
	\centering
	{\scriptsize{
    \begin{tabular}{|c|c|c|c|c|c|c|c|c|c|c|}
    \hline

    \textbf{}
    & \textbf{Discovery}
    & \multicolumn{3}{c|}{\textbf{Accuracy}}
    & \textbf{Gen.}
    & \multicolumn{3}{c|}{\textbf{Complexity}}
    & 
    & \textbf{Exec.}
    \\\cline{3-5}\cline{7-9}
    \textbf{Log}
    & \textbf{Method}
    & \textbf{Fitness}
    & \textbf{Precision}
    & \textbf{F-score}
    & \textbf{(3-Fold)}
    & \textbf{Size}
    & \textbf{CFC}
    & \textbf{Struct.}
    & \textbf{Sound}
    & \textbf{Time(s)} \\\hline

	%second row
	& $\alpha$\$ 
	& t/o & t/o & t/o & t/o
	& t/o & t/o & t/o & t/o & t/o
	\\
	
	%first row
	& IM 
	& \textbf{0.98}	& 0.50	& 0.66	& \textbf{0.98}
	& 59	& 37	& \textbf{1.00}	& yes	& 6.6
	\\
	
	& ETM 
	& 0.44 & \textbf{0.82} & 0.57 & t/o	
	& 67	& \textbf{16}		& \textbf{1.00}	& yes	& 14,400
	\\
	
BPIC12
	& FO 
	& -	& - & - & -
	& 102	& 117	& 0.13	& no	& 9.66
	\\
	
	%third row
%	& HM\textsubscript{6} 
%	& - & - & - & -  
%	& 85	& 99	& 0.05	& no	& \textbf{2.5}
%	\\
	
	%third row
	& S-HM\textsubscript{6} 
	& - & - & - & - 
	& 88 & 46 &  0.40 & no & 227.8
	\\
	
	%fourth row
	& HILP
	& - & - & - & -  
	& 300	& 460	& -	& no	& 772.2
	\\
	
	& SM
	& 0.75	& 0.76	& \textbf{0.76}	& 0.75
	& \textbf{53}	& 32	& 0.72	& yes	& \textbf{0.58}
	\\\cline{1-11}
	\hline

	%second row
	& $\alpha$\$ 
	& - & - & - & - 
	& 18 & 9 & - & no & 10,112.60
	\\
	
	%first row
	& IM 
	& 0.82	& \textbf{1.00}	& 0.90	& 0.82
	& \textbf{9} & 4 & \textbf{1.00} & yes	& 0.1
	\\
	
	& ETM 
	& \textbf{1.00}	& 0.70 & 0.82 & t/o
	& 38 & 38		& \textbf{1.00}	& yes	& 14,400
	\\
	
BPIC13\textsubscript{cp}
	& FO 
	& -	& - & - & -
	& 25	& 23	& 0.60	& no	& 0.06
	\\
	
	%third row
%	& HM\textsubscript{6} 
%	& - & - & - & -  
%	& 12 & 6 & 0.67	& no	& 	\textbf{0.1}
%	\\
	
	%third row
	& S-HM\textsubscript{6} 
	& 0.94	& 	0.99	& 	\textbf{0.97}	& 	\textbf{0.94}	
	& 15 & 6	& \textbf{1.00} & yes & 130.0
	\\

	%fourth row
	& HILP 
	& - & - & - & -
	& 	10	& 	\textbf{3}	& -	& yes	& 0.1
	\\
	
	& SM
	& 0.94	& 0.97	& 0.96	& \textbf{0.94}
	& 12 &	7	& \textbf{1.00}	& yes	& \textbf{0.03}
	\\\cline{1-11}
	\hline
	
	%second row
	& $\alpha$\$ 
	& 0.35 & 0.91 & 0.51 & t/o 
	& 15 & 7 & 0.47 & yes & 4,243.14
	\\
	
	%first row
	& IM 
	& 0.92	& 0.54	& 0.68	& \textbf{0.92}
	& 13 & 7	& \textbf{1.00}	& yes	& 1.0
	\\
	
	& ETM 
	& \textbf{1.00}	& 0.51 & 0.68 & t/o
	& 32	& 144		& \textbf{1.00}	& yes	& 14,400
	\\
	
	BPIC13\textsubscript{inc}
	& FO 
	& -	& - & - & -
	& 43	& 54	& 0.77	& no	& 1.41
	\\
	
	%third row
%	& HM\textsubscript{6} 
%	& 0.91 & \textbf{0.96} & \textbf{0.93} & 0.91 
%	& \textbf{9}	& \textbf{4}	& \textbf{1.00}	& yes	& \textbf{0.8}
%	\\
	
	%third row
	& S-HM\textsubscript{6} 
	& 0.91 & 0.96 & 0.93 & 0.91 
	& \textbf{9} & \textbf{4} & \textbf{1.00} & yes & 0.8
	\\

	%fourth row
	& HILP 
	& - & - & - & -
	& 24	& 9	& -	& yes	& 2.5
	\\
	
	& SM
	& 0.91	& \textbf{0.98} &	\textbf{0.94}	& 0.91
	& 13 & 9	& \textbf{1.00}	& yes	& \textbf{0.23}
	\\\cline{1-11}
	\hline

	& $\alpha$\$ 
	& 0.47 & 0.63 & 0.54 & t/o
	& 62 & 36 & 0.31 & yes & 14,057.48
	\\
	
	%first row
	& IM 
	& \textbf{0.89}	& 0.64 &	0.74	& \textbf{0.89}	
	& 31	& 18	& \textbf{1.00}	& yes	& 3.4
	\\
	
	& ETM 
	& 0.61	& \textbf{1.00} & \textbf{0.76} & t/o
	& \textbf{23}	& \textbf{9}		& \textbf{1.00}	& yes	& 14,400
	\\
	
	BPIC14\textsubscript{f}
	& FO 
	& -	& - & - & -
	& 37	& 46	& 0.38	& no	& 27.73
	\\
	
	%third row
%	& HM\textsubscript{6} 
%	& - & - & - & -
%	& 43 & 51 & - & no & \textbf{3.3}
%	\\
	
	%third row
	& S-HM\textsubscript{6} 
	& - & - & - & -
	& 202 & 132 & 0.73 & no	& 147.4
	\\
	
	%fourth row
	& HILP 
	& - & - & - & -	
	& 80 & 59 & - & no & 7.3
	\\
	
	& SM
	& 0.76	& 0.67	& 0.71	& 0.76
	& 27 & 16	& 0.74	& yes	& \textbf{0.59}
	\\\cline{1-11}
	\hline
	
		%second row
	& $\alpha$\$ 
	& 0.71 & 0.76 & 0.73 & t/o
	& 219 & 91 & 0.22 & yes & 3,545.9
	\\
	
	%first row
	& IM 
	& 0.97	& 0.57 &	0.71 &	\textbf{0.96} 
	&	164	& 108 &	\textbf{1.00} &	yes	& 0.6
	\\
	
	& ETM 
	& 0.56	& \textbf{0.94} &	0.70  & t/o
	& \textbf{67}	& \textbf{19}		& \textbf{1.00}	& yes	& 14,400
	\\
	
	BPIC15\textsubscript{1f}
	& FO 
	& \textbf{1.00}	& 0.76	& 0.87	& 0.94
	& 146	& 91	& 0.25	& yes	& 1.02
	\\
	
	%third row
%	& HM\textsubscript{6} 
%	& - & - & - & - 
%	& 150	& 98	& -	& no &	\textbf{0.5}
%	\\
	
	%third row
	& S-HM\textsubscript{6} 
	& -	& - & - & -
	& 204	& 116	& 0.56	& no	& 128.1
	\\
	
	%fourth row
	& HILP 
	& - & - & - & - 
	& 282	& 322 &	- &	no	& 4.4
	\\
	
	& SM
	& 0.90	& 0.88	& \textbf{0.89}	& 0.90
	& 110	& 43	& 0.50	& yes	& \textbf{0.48}
	\\\cline{1-11}
	\hline

	& $\alpha$\$
	& - & - & - & - 
	& 348 & 164 & 0.08 & no & 8,787.48
	\\
	
	%first row
	& IM 
	& 0.93	& 0.56	& 0.70	& 0.94
	& 193	& 123	& \textbf{1.00}	& yes	& 0.7
	\\
	
	& ETM 
	& 0.62 & \textbf{0.91} & 0.74 & t/o	
	& \textbf{95} & \textbf{32} & \textbf{1.00}	& yes	& 14,400
	\\
	
	BPIC15\textsubscript{2f}
	& FO 
	& -	& - & - & -
	& 195	& 159	& 0.09	& no	& 0.61
	\\
	
	%third row
%	& HM\textsubscript{6} 
%	& - & - & - & -  
%	& 194	& 158	& 0.11	& no	& \textbf{0.7}
%	\\
	
	%third row
	& S-HM\textsubscript{6} 
	& \textbf{0.98} & 0.59 & 0.74 & \textbf{0.97} 
	& 259	& 150	& 0.29	& yes	& 163.2
	\\
	
	%fourth row
	& HILP 
	& - & - & - & -	
	& - & - & - & - & t/o
	\\
	
	& SM
	& 0.77	& 0.90	& \textbf{0.83}	& 0.77
	& 122 & 41	& 0.32	& yes	& \textbf{0.25}
	\\\cline{1-11}
	\hline

	& $\alpha$\$
	& - & - & - & -
	& 319 & 169 & 0.03 & no & 10,118.15
	\\
	
	%first row
	& IM 
	& \textbf{0.95}	& 0.55	& 0.70	& \textbf{0.95}
	& 159	& 108	& \textbf{1.00}	& yes	& 1.3
	\\
	
	& ETM 
	& 0.68 & 0.88 & 0.76 & t/o
	& \textbf{84}	& \textbf{29}	& \textbf{1.00}	& yes	& 14,400
	\\
	
	BPIC15\textsubscript{3f}
	& FO 
	& -	& - & - & -
	& 174	& 164	& 0.06	& no	& 0.89
	\\
	
	%third row
%	& HM\textsubscript{6} 
%	& \textbf{0.95}	& 0.67	& \textbf{0.79}	& \textbf{0.95}
%	&	157	& 151	& 0.07	& yes	& \textbf{0.8}
%	\\
	
	%third row
	& S-HM\textsubscript{6} 
	& \textbf{0.95} & 0.67 & 0.79 & \textbf{0.95}
	& 159	& 151	& 0.13	& yes	& 139.9
	\\
	
	%fourth row
	& HILP 
	& - & - & - & -	
	& 433	& 829	& -	& no	& 1,062.9
	\\
	
	& SM
	& 0.78	& \textbf{0.94}	& \textbf{0.85}	& 0.78
	& 90	& \textbf{29}	& 0.61	& yes	& \textbf{0.36}
	\\\cline{1-11}
	\hline

	& $\alpha$\$
	& - & - & - & -
	& 272 & 128 & 0.13 & no & 6,410.25
	\\
	
	%first row
	& IM 
	& 0.96	& 0.58	& 0.73	& 0.96
	& 162	& 111	& \textbf{1.00}	& yes	& 0.7
	\\
	
	& ETM 
	& 0.65	& \textbf{0.93} & 0.77 & t/o
	& \textbf{83}	& \textbf{28}	& \textbf{1.00}	& yes	& 14,400
	\\
	
	BPIC15\textsubscript{4f}
	& FO 
	& -	& - & - & -
	& 157	& 127	& 0.14	& no	& 0.50
	\\
	
	%third row
%	& HM\textsubscript{6} 
%	& - & - & - & -
%	&	156	& 127	& 0.13 &	no	& \textbf{0.5}
%	\\
	
	%third row
	& S-HM\textsubscript{6} 
	& \textbf{0.99} & 0.64 & 0.78 & \textbf{0.99}
	& 209	& 137	& 0.37	& yes	& 136.9
	\\
	
	%fourth row
	& HILP 
	& - & - & - & -	
	& 364	& 593	& -	& no	& 14.7
	\\
	
	& SM
	& 0.73	& 0.91	& \textbf{0.81}	& 0.73
	& 96 & 31	& 0.31	& yes	& \textbf{0.25}
	\\\cline{1-11}
	\hline

	& $\alpha$\$
	& 0.62 & 0.75 & 0.68 & t/o
	& 280 & 126 & 0.10 & yes & 7,603.19
	\\
	
	%first row
	& IM 
	& 0.94	& 0.18	& 0.30	& 0.94
	& 134	& 95	& \textbf{1.00}	& yes	& 1.5
	\\
	
	& ETM 
	& 0.57	& \textbf{0.94} &	0.71 & t/o
	& \textbf{88}	& \textbf{18}		& \textbf{1.00}	& yes	& 14,400
	\\
	
	BPIC15\textsubscript{5f}
	& FO 
	& \textbf{1.00}	& 0.71	& 0.83	& \textbf{1.00}
	& 166	& 125	& 0.15	& yes	& 0.56
	\\
	
	%third row
%	& HM\textsubscript{6} 
%	& - & - & - & -
%	& 166	& 124	& 0.15 &	no	& \textbf{1.2}
%	\\
	
	%third row
	& S-HM\textsubscript{6} 
	& \textbf{1.00}	& 0.70	& 0.82	& \textbf{1.00}
	& 211	& 135	& 0.35	& yes	& 141.9
	\\
	
	%fourth row
	& HILP 
	& - & - & - & -
	& - & - & - & - & t/o
	\\
	
	& SM
	& 0.79	& \textbf{0.94}	& \textbf{0.86}	& 0.79
	& 102	& 30	& 0.33	& yes	& \textbf{0.27}
	\\\cline{1-11}
	\hline

	& $\alpha$\$
	& t/o & t/o & t/o & t/o
	& t/o & t/o & t/o & t/o & t/o
	\\
	
	%first row
	& IM 
	& \textbf{0.98}	& 0.70	& 0.82	& \textbf{0.98}
	& 35 &	20 &	\textbf{1.00} &	yes &	13.3
	\\
	
	& ETM 
	& 0.76	& \textbf{1.00}	& 0.86 & t/o
	& 42	& \textbf{4} & \textbf{1.00}	& yes	& 14,400
	\\
	
	BPIC17\textsubscript{f}
	& FO 
	& -	& - & - & -
	& 98	& 82	& 0.25	& no	& 64.33
	\\
	
	%third row
%	& HM\textsubscript{6} 
%	& - & - & - & -
%	& \textbf{29} &	10 &	0.45 &	no &	\textbf{6.5}
%	\\
	
	%third row
	& S-HM\textsubscript{6} 
	& 0.95 & 0.62 & 0.75 & 0.94
	& 42	& 13	& 0.97	& yes	& 143.2
	\\
	
	%fourth row
	& HILP 
	& - & - & - & -	
	& 222 &	330	& - & no	& 384.5
	\\
	
	& SM
	& 0.95	& 0.85	& \textbf{0.90}	& 0.95
	& \textbf{32} & 17	& 0.75	& yes	& \textbf{2.53}
	\\\cline{1-11}
	\hline
	
	\end{tabular}
	}}
  	\caption{Default parameters evaluation results for the BPIC logs.}\label{tab:publogs1}
\end{table*}

\begin{table*}[tbp]
	\centering
	{\scriptsize{
    \begin{tabular}{|c|c|c|c|c|c|c|c|c|c|c|}
    \hline

    \textbf{}
    & \textbf{Discovery}
    & \multicolumn{3}{c|}{\textbf{Accuracy}}
    & \textbf{Gen.}
    & \multicolumn{3}{c|}{\textbf{Complexity}}
    & 
    & \textbf{Exec.}
    \\\cline{3-5}\cline{7-9}
    \textbf{Log}
    & \textbf{Method}
    & \textbf{Fitness}
    & \textbf{Precision}
    & \textbf{F-score}
    & \textbf{(3-Fold)}
    & \textbf{Size}
    & \textbf{CFC}
    & \textbf{Struct.}
    & \textbf{Sound?}
    & \textbf{Time (sec)} \\\hline

	%second row
	& $\alpha$\$
	& t/o & t/o & t/o & t/o
	& t/o & t/o & t/o & t/o & t/o
	\\
	
	%first row
	& IM 
	& 0.99	& 0.70	& 0.82	& 0.99	
	& 34	& 20	& \textbf{1.00}	& yes	& 10.9
	\\
	
	& ETM 
	& 0.99	& 0.92	& 0.95 & t/o
	& 57	& 32 & \textbf{1.00}	& yes	& 14,400
	\\
	
	RTFMP
	& FO 
	& \textbf{1.00}	& 0.94	& 0.97	& 0.97
	& 31	& 32	& 0.19	& yes	& 2.57
	\\
	
	%third row
%	& HM\textsubscript{6} 
%	& - & - & - & -
%	& 47	& 50	& 0.06	& no	& 7.8
%	\\
	
	%third row
	& S-HM\textsubscript{6} 
	& 0.98 & 0.95 & 0.96 & 0.98 
	& 163 & 97 & \textbf{1.00} & yes & 262.7
	\\
	
	%fourth row
	& HILP 
	& - & - & - & -
	& 57	& 53	& -	& no	& 3.5
	\\
	
	& SM
	& 0.99	& \textbf{1.00}	& \textbf{1.00}	& \textbf{1.00}
	& \textbf{22} & \textbf{16}	& 0.46	& yes	& \textbf{1.25}
	\\\cline{1-11}
	\hline

	%second row
	& $\alpha$\$
	& - & - & - & -
	& 146 & 156 & 0.01 & no & 3,883.12
	\\
	
	%first row
	& IM 
	& \textbf{0.99}	& 0.45	& 0.62	& \textbf{0.96}	
	& 50	& 32	& \textbf{1.00}	& yes	& 0.4
	\\
	
	& ETM 
	& 0.83	& 0.66	& 0.74 & t/o
	& 108	& 101		& \textbf{1.00}	& yes	& 14,400
	\\
	
	SEPSIS
	& FO 
	& -	& - & - & -
	& 60	& 63 & 0.28	& no	& 0.17
	\\
	
	%third row
%	& HM\textsubscript{6} 
%	& - & - & - & -
%	& 81	& 132	& 0.17	& no	& \textbf{0.03}
%	\\
	
	%third row
	& S-HM\textsubscript{6} 
	& 0.92 & 0.42 & 0.58 & 0.92
	& 279 & 198 &  \textbf{1.00} & yes & 242.7
	\\
	
	%fourth row
	& HILP 
	& - & - & - & -
	& 87 & 129 & - & no	& 1.6
	\\
	
	& SM
	& 0.73	& \textbf{0.86}	& \textbf{0.79}	& 0.73
	& \textbf{31} & \textbf{20}	& 0.97	& yes	& \textbf{0.05}
	\\\cline{1-11}
	\hline

	\end{tabular}
    }}
  	\caption{Default parameters evaluation results for the public logs.}\label{tab:publogs2}
\end{table*}
%\vspace{-\baselineskip}
% !TEX root = main.tex
\begin{table*}
	\centering
	{\scriptsize{
    \begin{tabular}{|c|c|c|c|c|c|c|c|c|c|c|}
    \hline

    \textbf{}
    & \textbf{Discovery}
    & \multicolumn{3}{c|}{\textbf{Accuracy}}
    & \textbf{Gen.}
    & \multicolumn{3}{c|}{\textbf{Complexity}}
    & 
    & \textbf{Exec.}
    \\\cline{3-5}\cline{7-9}
    \textbf{Log}
    & \textbf{Method}
    & \textbf{Fitness}
    & \textbf{Precision}
    & \textbf{F-score}
    & \textbf{(3-Fold)}
    & \textbf{Size}
    & \textbf{CFC}
    & \textbf{Struct.}
    & \textbf{Sound}
    & \textbf{Time(s)} \\\hline

	%second row
	& $\alpha$\$ 
	& - & - & - & t/o
	& 45 & 34 & - & no & 11,168.54
	\\
	
	%first row
	& IM 
	& 0.90	& 0.67	& 0.77	& 0.90	
	& \textbf{20}	& \textbf{9}	& \textbf{1.00}	& yes	&2.08
	\\
	
	& ETM 
	& \textbf{0.99}	& 0.81	& 0.89	& t/o	
	& 23	& 12	& \textbf{1.00}	& yes	& 14,400
	\\
	
	PRT1
	& FO 
	& -	& -	& -	& -	
	& 30	& 28	& 0.53	& no	& 0.95
	\\
	
	%third row
%	& HM\textsubscript{6} 
%	& - & - & - & -  
%	& 85	& 99	& 0.05	& no	& \textbf{2.5}
%	\\
	
	%third row
	& S-HM\textsubscript{6} 
	& 0.88	& 0.77	& 0.82	& 0.88	
	& 59	& 39	& \textbf{1.00}	& yes	& 122.16
	\\
	
	%fourth row
	& HILP
	& - & - & - & -
	& 195	& 271	& -	& no	& 2.59
	\\
	
	& SM 
	& 0.98	& \textbf{0.99}	& \textbf{0.98}		& \textbf{0.98}		
	& 29	& 18	& 0.79	& yes	& \textbf{0.47}
	\\\cline{1-11}
	\hline

	%second row
	& $\alpha$\$ 
	& - & - & - & - 
	& 134 & 113 & 0.25 & no & 3,438.72
	\\
	
	%first row
	& IM 
	& ex & ex & ex & ex
	& 45 & 33	& \textbf{1.00}	& yes	& 1.41
	\\
	
	& ETM 
	& 0.57	& \textbf{0.94}	& 0.71	& t/o	
	& 86	& \textbf{21}	& \textbf{1.00}	& yes	& 14,400
	\\
	
	PRT2
	& FO 
	& - & - & - & -
	& 76 & 74	& 0.59	& no	& 0.88
	\\
	
	%third row
%	& HM\textsubscript{6} 
%	& - & - & - & -  
%	& 12 & 6 & 0.67	& no	& 	\textbf{0.1}
%	\\
	
	%third row
	& S-HM\textsubscript{6} 
	& -	& -	& -	& -	
	& 67	& 105	& 0.43	& no	& 1.77
	\\

	%fourth row
	& HILP 
	& - & - & - & -
	& 190	& 299	& -	& no	& 21.33
	\\
	
	& SM 
	& \textbf{0.81}	& 0.74	& \textbf{0.77}	& \textbf{0.81}	
	& \textbf{40}	& 30	& 0.85	& yes	& \textbf{0.31}
	\\\cline{1-11}
	\hline
	
	%second row
	& $\alpha$\$ 
	& 0.67	& 0.76	& 0.71	& 0.67	
	& 70	& 40	& 0.11	& yes	& 220.11
	\\
	
	%first row
	& IM 
	& 0.98	& 0.68	& 0.80	& 0.98	
	& 37	& 20	& \textbf{1.00}	& yes	& 0.44
	\\
	
	& ETM 
	& 0.98	& 0.86	& \textbf{0.92}	& t/o	
	& 51	& 37	& \textbf{1.00}	& yes	& 14,400
	\\
	
	PRT3
	& FO 
	& \textbf{1.00}	& 0.86	& \textbf{0.92}	& \textbf{1.00}	
	& 34	& 37	& 0.32	& yes	& 0.50
	\\
	
	%third row
%	& HM\textsubscript{6} 
%	& 0.91 & \textbf{0.96} & \textbf{0.93} & 0.91 
%	& \textbf{9}	& \textbf{4}	& \textbf{\textbf{1.00}}	& yes	& \textbf{0.8}
%	\\
	
	%third row
	& S-HM\textsubscript{6} 
	& \textbf{1.00}	& 0.83	& 0.91	& \textbf{1.00}	
	& 40	& 38	& 0.43	& yes	& 0.67
	\\

	%fourth row
	& HILP 
	& - & - & - & -
	& 343	& 525	& -	& no	& 0.73
	\\
	
	& SM 
	& 0.83	& \textbf{0.91}	& 0.87	& 0.88	
	& \textbf{32}	& \textbf{17}	& 0.78	& yes	& \textbf{0.17}
	\\\cline{1-11}
	\hline

	& $\alpha$\$ 
	& 0.86	& 0.93	& 0.90	& t/o	
	& \textbf{21}	& \textbf{10}	& \textbf{1.00}	& yes &	13,586.48
	\\
	
	%first row
	& IM 
	& 0.93	& 0.75	& 0.83	& 0.93	
	& 27	& 13	& \textbf{1.00}	& yes	& 1.33
	\\
	
	& ETM 
	& 0.84	& 0.85	& 0.84	& t/o	
	& 64	& 28	& \textbf{1.00}	& yes	& 14,400
	\\
	
	PRT4
	& FO 
	& - & - & - & -
	& 37 & 40	& 0.54	& no	& 6.33
	\\
	
	%third row
%	& HM\textsubscript{6} 
%	& - & - & - & -
%	& 43 & 51 & - & no & \textbf{3.3}
%	\\
	
	%third row
	& S-HM\textsubscript{6} 
	& \textbf{1.00}	& 0.86	& \textbf{0.93}	& \textbf{1.00}	
	& 370	& 274	& \textbf{1.00}	& yes	& 241.57
	\\
	
	%fourth row
	& HILP 
	& - & - & - & -
	& 213	& 306	& -	& no	& 5.31
	\\
	
	& SM 
	& 0.87	& \textbf{0.99}	& \textbf{0.93}	& 0.85	
	& 34	& 21	& 0.65	& yes	& \textbf{0.45}
	\\\cline{1-11}
	\hline

		%second row
	& $\alpha$\$ 
	& 1.00	& 1.00	& 1.00	& 1.00	
	& 10	& 1	& 1.00	& yes & 2.02
	\\
	
	%first row
	& IM 
	& 1.00	& 1.00	& 1.00	& 1.00	
	& 10	& 1	& 1.00	& yes	& 0.03
	\\
	
	& ETM 
	& 1.00	& 1.00	& 1.00	& 1.00	
	& 10	& 1	& 1.00	& yes	& 2.49
	\\
	
	PRT5
	& FO 
	& 1.00	& 1.00	& 1.00	& 1.00	
	& 10	& 1	& 1.00	& yes	& \textbf{0.02}
	\\
	
	%third row
%	& HM\textsubscript{6} 
%	& - & - & - & - 
%	& 150	& 98	& -	& no &	\textbf{0.5}
%	\\
	
	%third row
	& S-HM\textsubscript{6} 
	& 1.00	& 1.00	& 1.00	& 1.00	
	& 10	& 1	& 1.00	& yes	 & 0.11
	\\
	
	%fourth row
	& HILP 
	& 1.00	& 1.00	& 1.00	& 1.00	
	& 10	& 1	& 1.00	& yes	& 0.05
	\\
	
	& SM 
	& 1.00	& 1.00	& 1.00	& 1.00	
	& 10	& 1	& 1.00	& yes	& \textbf{0.02}
	\\\cline{1-11}
	\hline

	& $\alpha$\$
	& 0.80	& 0.77	& 0.79	& 0.80	
	& 38	& 17	& 0.24	& yes	& 40.10
	\\
	
	%first row
	& IM 
	& 0.99	& 0.82	& 0.90	& 0.99	
	& 23	& 10	& \textbf{1.00}	& yes	& 2.30
	\\
	
	& ETM 
	& 0.98	& 0.80	& 0.88	& t/o	
	& 41	& 16	& \textbf{1.00}	& yes	& 14,400
	\\
	
	PRT6
	& FO 
	& \textbf{1.00}	& 0.91	& 0.95	& \textbf{1.00}	
	& 22	& 17	& 0.41	& yes	& 0.05
	\\
	
	%third row
%	& HM\textsubscript{6} 
%	& - & - & - & -  
%	& 194	& 158	& 0.11	& no	& \textbf{0.7}
%	\\
	
	%third row
	& S-HM\textsubscript{6} 
	& \textbf{1.00}	& 0.91	& 0.95	& \textbf{1.00}	
	& 22	& 17	& 0.41	& yes	& 0.42
	\\
	
	%fourth row
	& HILP 
	& - & - & - & -
	& 157	& 214	& -	& no	& 0.13
	\\
	
	& SM 
	& 0.94	& \textbf{1.00}	& \textbf{0.97}	& 0.94	
	& \textbf{16}	& \textbf{5}	& \textbf{1.00}	& yes	& \textbf{0.02}
	\\\cline{1-11}
	\hline
	
		\end{tabular}
	}}
  	\caption{Default parameters evaluation results for the proprietary logs - 1.}\label{tab:prtlogs1}
\end{table*}

\begin{table*}[tbp]
	\centering
	{\scriptsize{
    \begin{tabular}{|c|c|c|c|c|c|c|c|c|c|c|}
    \hline

    \textbf{}
    & \textbf{Discovery}
    & \multicolumn{3}{c|}{\textbf{Accuracy}}
    & \textbf{Gen.}
    & \multicolumn{3}{c|}{\textbf{Complexity}}
    & 
    & \textbf{Exec.}
    \\\cline{3-5}\cline{7-9}
    \textbf{Log}
    & \textbf{Method}
    & \textbf{Fitness}
    & \textbf{Precision}
    & \textbf{F-score}
    & \textbf{(3-Fold)}
    & \textbf{Size}
    & \textbf{CFC}
    & \textbf{Struct.}
    & \textbf{Sound?}
    & \textbf{Time (sec)} \\\hline
	
	& $\alpha$\$
	& 0.85	& 0.90	& 0.88	& 0.85	
	& 29	& \textbf{9}	& 0.48	& yes	& 143.66
	\\
	
	%first row
	& IM 
	& \textbf{1.00}	& 0.73	& 0.84	& \textbf{1.00}	
	& 29 &	13	& \textbf{1.00}	& yes	& 0.13
	\\
	
	& ETM 
	& 0.90	& 0.81	& 0.85	& t/o	
	& 60	& 29	& \textbf{1.00}	& yes	& 14,400
	\\
	
	PRT7
	& FO 
	& 0.99	& \textbf{1.00}	& 0.99	& 0.99	
	& \textbf{26}	& 16	& 0.39	& yes	& 0.08
	\\
	
	%third row
%	& HM\textsubscript{6} 
%	& \textbf{0.95}	& 0.67	& \textbf{0.79}	& \textbf{0.95}
%	&	157	& 151	& 0.07	& yes	& \textbf{0.8}
%	\\
	
	%third row
	& S-HM\textsubscript{6} 
	& \textbf{1.00}	& \textbf{1.00}	& \textbf{1.00}	& \textbf{1.00}	
	& 163	& 76	& \textbf{1.00}	& yes	& 249.74
	\\
	
	%fourth row
	& HILP 
	& - & - & - & -
	& 278	& 355	& -	& no	& 0.27
	\\
	
	& SM 
	& 0.91	& \textbf{1.00}	& 0.95	& 0.92	
	& 30	& 11	& 0.33	& yes	& \textbf{0.06}
	\\\cline{1-11}
	\hline

	& $\alpha$\$
	& t/o & t/o & t/o & t/o
	& t/o & t/o & t/o & t/o & t/o
	\\
	
	%first row
	& IM 
	& \textbf{0.98}	& 0.33	& 0.49	& \textbf{0.93}	
	& 111	& 92	& \textbf{1.00}	& yes	& \textbf{0.41}
	\\
	
	& ETM 
	& 0.35	& \textbf{0.88}	& 0.50 & t/o
	& \textbf{75} & \textbf{12} & \textbf{1.00} & yes & 14,400
	\\
	
	PRT8
	& FO 
	& - & - & - & -
	& 228	& 179	& 0.74	& no	& 0.55
	\\
	
	%third row
%	& HM\textsubscript{6} 
%	& - & - & - & -
%	&	156	& 127	& 0.13 &	no	& \textbf{0.5}
%	\\
	
	%third row
	& S-HM\textsubscript{6} 
	& - & - & - & -
	& 388	& 323	& 0.87	& no	& 370.66
	\\
	
	%fourth row
	& HILP 
	& t/o & t/o & t/o & t/o
	& t/o & t/o & t/o & t/o & t/o
	\\
	
	& SM 
	& 0.97	& 0.67	& \textbf{0.79}	& 0.92	
	& 406	& 488	& 0.43	& yes	& 1.28
	\\\cline{1-11}
	\hline

	& $\alpha$\$
	& t/o & t/o & t/o & t/o
	& t/o & t/o & t/o & t/o & t/o
	\\
	
	%first row
	& IM 
	& 0.90	& 0.61	& 0.73	& 0.89	
	& 28	& 16	& \textbf{1.00}	& yes	& 63.70
	\\
	
	& ETM 
	& 0.75	& 0.49	& 0.59	& 0.74	
	& \textbf{27}	& \textbf{13}	& \textbf{1.00}	& yes	& 1,266.71
	\\
	
	PRT9
	& FO 
	& - & - & - & -
	& 32 &	45	& 0.72	& no	& 42.83
	\\
	
	%third row
%	& HM\textsubscript{6} 
%	& - & - & - & -
%	& 166	& 124	& 0.15 &	no	& \textbf{1.2}
%	\\
	
	%third row
	& S-HM\textsubscript{6} 
	& \textbf{0.96}	& 0.98	& \textbf{0.97}	& \textbf{0.96}	
	& 723	& 558	& \textbf{1.00}	& yes	& 318.69
	\\
	
	%fourth row
	& HILP 
	& - & - & - & -
	& 164	& 257	& -	& no	& 51.47
	\\
	
	& SM 
	& 0.92	& \textbf{1.00}	& 0.96	& 0.92	
	& 30	& 20	& \textbf{1.00}	& yes	& \textbf{9.11}
	\\\cline{1-11}
	\hline
	
	& $\alpha$\$
	& t/o & t/o & t/o & t/o
	& t/o & t/o & t/o & t/o & t/o
	\\
	
	%first row
	& IM 
	& 0.96	& 0.79	& 0.87	& 0.96	
	& \textbf{41}	& \textbf{29}	& \textbf{1.00}	& yes	& 2.50
	\\
	
	& ETM 
	& \textbf{1.00}	& 0.63	& 0.77	& t/o	
	& 61	& 45	& \textbf{1.00}	& yes	& 14,400
	\\
	
	PRT10
	& FO 
	& 0.99	& 0.93	& 0.96	& \textbf{0.99}	
	& 52	& 85	& 0.64	& yes	& 0.98
	\\
	
	%third row
%	& HM\textsubscript{6} 
%	& - & - & - & -
%	& \textbf{29} &	10 &	0.45 &	no &	\textbf{6.5}
%	\\
	
	%third row
	& S-HM\textsubscript{6} 
	& - & - & - & -
	& 77	& 110	& -	& no &	1.81
	\\
	
	%fourth row
	& HILP 
	& - & - & - & -
	& 846	& 3130	& -	& no	& 2.55
	\\
	
	& SM 
	& 0.97	& \textbf{0.97}	& \textbf{0.97}	& 0.97	
	& 79	& 68	& 0.43	& yes	& \textbf{0.47}
	\\\cline{1-11}
	\hline
	
	%second row
	& $\alpha$\$
	& t/o & t/o & t/o & t/o
	& t/o & t/o & t/o & t/o & t/o
	\\
	
	%first row
	& IM 
	& t/o	& t/o	& t/o	& t/o	
	& 549	& 365	& \textbf{1.00}	& yes	& 121.50
	\\
	
	& ETM 
	& \textbf{0.10}	& \textbf{1.00}	& \textbf{0.18}	& t/o	
	& \textbf{21} &	\textbf{3}	& \textbf{1.00}	& yes	& 14,400

	\\
	
	PRT11
	& FO 
	& - & - & - & -
	& 680 & 713	& 0.68	& no	& 81.33
	\\
	
	%third row
%	& HM\textsubscript{6} 
%	& - & - & - & -
%	& 47	& 50	& 0.06	& no	& 7.8
%	\\
	
	%third row
	& S-HM\textsubscript{6} 
	& ex & ex & ex & ex
	& ex & ex & ex & ex & ex
	\\
	
	%fourth row
	& HILP 
	& t/o & t/o & t/o & t/o
	& t/o & t/o & t/o & t/o & t/o
	\\
	
	& SM 
	& - & - & - & -
	& 712	& 609	& 0.12	& no	& \textbf{19.53}
	\\\cline{1-11}
	\hline
	
	%second row
	& $\alpha$\$
	& t/o & t/o & t/o & t/o
	& t/o & t/o & t/o & t/o & t/o
	\\
	
	%first row
	& IM 
	& \textbf{1.00}	& 0.77	& 0.87	& \textbf{1.00}	
	& 32	& 25	& \textbf{1.00}	& yes	&3.94
	\\
	
	& ETM 
	& 0.63	& \textbf{1.00}	& 0.77	& t/o	
	& \textbf{21}	& \textbf{8}	& \textbf{1.00}	& yes	& 14,400
	\\
	
	PRT12
	& FO 
	& - & - & - & -
	& 87 &	129	& 0.38	& no	& 1.67
	\\
	
	%third row
%	& HM\textsubscript{6} 
%	& - & - & - & -
%	& 81	& 132	& 0.17	& no	& \textbf{0.03}
%	\\
	
	%third row
	& S-HM\textsubscript{6} 
	& -	& -	& -	& -	
	& 4370	& 3191	& \textbf{1.00}	& yes	& 347.57
	\\
	
	%fourth row
	& HILP 
	& - & - & - & -
	& 926	& 2492	& -	& no	& 7.34
	\\
	
	& SM 
	& 0.96	& 0.99	& \textbf{0.97}	& 0.95	
	& 97	& 84	& 0.58	& yes	& \textbf{0.36}
	\\\cline{1-11}
	\hline

	\end{tabular}
    }}
  	\caption{Default parameters evaluation results for the proprietary logs - 2.}\label{tab:prtlogs2}
\end{table*}
%\vspace{-\baselineskip}
% !TEX root = main.tex
\begin{table}
	\centering
    \begin{tabular}{|c|c|c|c|c|c|}
    \hline

    \textbf{Log}
    & \textbf{Metrics}
    & \textbf{IM}
    & \textbf{FO}
    & \textbf{S-HM\textsubscript{6}}
    & \textbf{SM}
    \\\hline
	
%START-LOG	
	& Fitness
	& 0.81 & 0.82 & 0.96 & 0.91
	\\
	
BPIC12
	& Precision
	& 0.64 & 0.41 & 0.66 & 0.83
	\\
	
	& F-score	
	& \textit{0.71} & \textit{0.54} & \textit{0.78} & \textit{\textbf{0.87}}
    \\\hline
%END-LOG	

%START-LOG	
	& Fitness
	& 0.99 & - & 0.96 & 0.94
	\\
	
BPIC13\textsubscript{cp}
	& Precision
	& 0.98 & - & 1.00 & 0.97
	\\
	
	& F-score	
	& \textit{0.98} & - & \textit{\textbf{0.98}} & 0.96
    \\\hline
%END-LOG	

%START-LOG	
	& Fitness
	& 1.00 & - & 0.93 & 0.91
	\\
	
BPIC13\textsubscript{inc}
	& Precision
	& 0.71 & - & 0.98 & 0.98
	\\
	
	& F-score	
	& \textit{0.83} & - & \textit{\textbf{0.96}} & 0.94
    \\\hline
%END-LOG	

%START-LOG	
	& Fitness
	& 0.75 & 0.94 & 0.91 & 0.80
	\\
	
BPIC14\textsubscript{f}
	& Precision
	& 0.97 & 0.85 & 0.84 & 0.99
	\\
	
	& F-score	
	& \textit{0.85} & \textit{\textbf{0.89}} & \textit{0.88} & \textit{\textbf{0.89}}
    \\\hline
%END-LOG

%START-LOG	
	& Fitness
	& 1.00 & 0.90 & 0.88 & 0.95
	\\

BPIC15\textsubscript{1f}
	& Precision
	& 0.57 & 0.88 & 0.89 & 0.86
	\\
	
	& F-score	
	& \textit{0.72} & \textit{0.89} & \textit{0.89} & \textit{\textbf{0.90}}
    \\\hline
%END-LOG	

%START-LOG	
	& Fitness
	& 0.69 & 0.99 & 0.99 & 0.81
	\\
	
BPIC15\textsubscript{2f}
	& Precision
	& 0.79 & 0.63 & 0.62 & 0.86
	\\
	
	& F-score	
	& \textit{0.74} & \textit{0.77} & \textit{0.76} & \textbf{0.83}
    \\\hline
%END-LOG	

%START-LOG	
	& Fitness
	& 0.77 & 0.80 & 0.81 & 0.78
	\\
	
BPIC15\textsubscript{3f}
	& Precision
	& 0.80 & 0.86 & 0.77 & 0.94
	\\
	
	& F-score	
	& \textit{0.79} & \textit{0.83} & 0.79 & \textbf{0.85}
    \\\hline
%END-LOG	

%START-LOG	
	& Fitness
	& 0.73 & 0.76 & 0.99 & 0.77
	\\
	
BPIC15\textsubscript{4f}
	& Precision
	& 0.87 & 0.87 & 0.66 & 0.90
	\\
	
	& F-score	
	& \textit{0.80} & \textit{0.81} & \textit{0.79} & \textit{\textbf{0.83}}
    \\\hline
%END-LOG	

%START-LOG	
	& Fitness
	& 0.65 & 0.81 & 0.82 & 0.86
	\\
	
BPIC15\textsubscript{5f}
	& Precision
	& 0.87 & 0.91 & 0.94 & 0.90
	\\
	
	& F-score	
	& \textit{0.75} & \textit{0.86} & \textit{0.87} & \textit{\textbf{0.88}}
    \\\hline
%END-LOG	

%START-LOG	
	& Fitness
	& 1.00 & 1.00 & 0.97 & 0.95
	\\
	
BPIC17\textsubscript{f}
	& Precision
	& 0.70 & 0.07 & 0.70 & 0.85
	\\
	
	& F-score	
	& 0.82 & \textit{0.12} & \textit{0.81} & \textbf{0.90}
    \\\hline
%END-LOG	

%START-LOG	
	& Fitness
	& 0.96 & 1.00 & 0.98 & 0.99
	\\
	
RTFMP
	& Precision
	& 0.72 & 0.94 & 0.96 & 1.00
	\\
	
	& F-score	
	& 0.82 & 0.97 & \textit{0.97} & \textbf{1.00}
    \\\hline
%END-LOG	

%START-LOG	
	& Fitness
	& 0.66 & 0.74 & 0.92 & 0.85
	\\
	
SEPSIS
	& Precision
	& 0.91 & 0.67 & 0.42 & 0.73
	\\
	
	& F-score	
	& \textit{0.76} & \textit{0.70} & 0.58 & \textbf{0.79}
    \\\hline
%END-LOG	
				%% PROPRIETARY LOGS
%START-LOG	
	& Fitness
	& 1.00 & 0.98 & 0.96 & 0.98
	\\
		
PRT1
	& Precision
	& 0.82 & 0.92 & 0.98 & 0.99
	\\
	
	& F-score	
	& \textit{0.90} & \textit{0.95} & \textit{0.97} & \textbf{0.98}
    \\\hline
%END-LOG	

%START-LOG	
	& Fitness
	& - & 1.00 & - & 0.81
	\\
	
PRT2
	& Precision
	& - & 0.17 & - & 0.74
	\\
	
	& F-score	
	& -	& \textit{0.30} & - & \textbf{0.77}
    \\\hline
%END-LOG	

%START-LOG	
	& Fitness
	& 0.87 & 1.00 & 0.99 & 1.00
	\\
	
PRT3
	& Precision
	& 0.93 & 0.86 & 0.85 & 0.92
	\\
	
	& F-score	
	& \textit{0.90} & 0.92 & 0.91 & \textit{\textbf{0.96}}
    \\\hline
%END-LOG	

%START-LOG	
	& Fitness
	& 0.86 & 1.00 & 0.93 & 0.99
	\\
	
PRT4
	& Precision
	& 1.00 & 0.87 & 0.96 & 1.00
	\\
	
	& F-score	
	& \textit{0.92} & \textit{0.93} & \textit{0.95} & \textit{\textbf{0.99}}
    \\\hline
%END-LOG	

%START-LOG	
	& Fitness
	& 1.00 & 1.00 & 1.00 & 1.00
	\\
	
PRT5
	& Precision
	& 1.00 & 1.00 & 1.00 & 1.00
	\\
	
	& F-score	
	& 1.00 & 1.00 & 1.00 & 1.00
    \\\hline
%END-LOG	

%START-LOG	
	& Fitness
	& 0.92 & 1.00 & 0.98 & 0.94
	\\
	
PRT6
	& Precision
	& 1.00 & 0.91 & 0.96 & 1.00
	\\
	
	& F-score	
	& \textit{0.96} & 0.95 & \textit{\textbf{0.97}} & \textbf{0.97}
    \\\hline
%END-LOG	

%START-LOG	
	& Fitness
	& 0.88 & 0.99 & 1.00 & 0.93
	\\
	
PRT7
	& Precision
	& 1.00 & 1.00 & 1.00 & 1.00
	\\
	
	& F-score	
	& \textit{0.93} & 0.99 & \textbf{1.00} & \textit{0.96}
    \\\hline
%END-LOG	

%START-LOG	
	& Fitness
	& 0.79 & 1.00 & 0.93 & 0.99
	\\
	
PRT8
	& Precision
	& 0.37 & 0.14 & 0.42 & 0.66
	\\
	
	& F-score	
	& \textit{0.51} & \textit{0.25} & \textit{0.58} & \textbf{0.79}
    \\\hline
%END-LOG	

%START-LOG	
	& Fitness
	& 0.93 & - & 0.96 & 0.99
	\\
	
PRT9
	& Precision
	& 0.68 & - & 0.98 & 1.00
	\\
	
	& F-score	
	& \textit{0.78} & - & 0.97 & \textit{\textbf{0.99}}
    \\\hline
%END-LOG	\\

%START-LOG	
	& Fitness
	& 0.99 & 0.99 & 0.98 & 0.99
	\\
	
PRT10
	& Precision
	& 0.79 & 0.93 & 0.83 & 0.97
	\\
	
	& F-score	
	& \textit{0.88} & 0.96 & \textit{0.90} & \textit{\textbf{0.98}}
    \\\hline
%END-LOG	

%START-LOG	
	& Fitness
	& ex & - & - & -
	\\
	
PRT11
	& Precision
	& ex & - & - & -
	\\
	
	& F-score	
	& ex & - & - & -
    \\\hline
%END-LOG	

%START-LOG	
	& Fitness
	& 0.97 & 1.00 & - & 0.98
	\\
	
PRT12
	& Precision
	& 0.85 & 0.80 & - & 0.97
	\\
	
	& F-score	
	& \textit{0.91} & \textit{0.89} & - & \textit{\textbf{0.98}}
    \\\hline
%END-LOG	
	
	\end{tabular}
  	\caption{Hyperparameters-optimization evaluation results.}\label{tab:hyperdata}
\end{table}

We performed two types of evaluations. The first evaluation was meant to compare all the process discovery methods using their default parameters. In the second evaluation, we wanted to perform a similar comparison using hyper-parameter optimization. Due to  their extremely long execution times, for this second evaluation we held out $\alpha\$$ and ETM which would have been prohibitive for a hyper-parameter optimization exercise. Additionally, we excluded HILP since we did not find any input parameters which could be used to optimize the f-score of the models produced. For the remaining four methods, we evaluated the following input parameters: the two filtering thresholds required in input by SM and S-HM\textsubscript{6}, the single threshold required in input by IM, and the threshold and the boolean flag required in input by FO. Since all the thresholds range from 0.0 to 1.0, we used steps of 0.05 for IM, and steps of 0.10 for the thresholds of SM, S-HM\textsubscript{6} and FO. For FO, we considered all the possible combinations of the filtering threshold and the boolean flag.\footnote{This to have a similar number of data points across all methods.}

%We performed two types of evaluations: a default-parameters evaluation and a hyper-parameter optimization evaluation. In the first evaluation, we compared all the process discovery methods using their default parameters. In the second evaluation, we compared only four out of the seven discovery methods. Precisely, we held out $\alpha\$$ and ETM because their execution times are prohibitive for a hyper-parameter optimization exercise. Whilst we excluded HILP since we did not find any input parameters to optimize the f-score of the models produced by this technique. For the remaining four methods, we evaluated the following input parameters: the two filtering thresholds required in input by SM and S-HM\textsubscript{6}, the single threshold required in input by IM, and the threshold and the boolean flag required in input by FO. Since all the thresholds range from 0.0 to 1.0,
%we used steps of 0.05 for IM, and steps of 0.10 for the thresholds of SM, S-HM\textsubscript{6} and FO. For FO, we considered all the possible combinations of the filtering threshold and the boolean flag.

The results of the first evaluation are shown in the tables~\ref{tab:publogs1},~\ref{tab:publogs2},~\ref{tab:prtlogs1}, and~\ref{tab:prtlogs2}, where the best score for each measure on each log is highlighted in bold. In the tables, we used ``-'' to report that a given accuracy or complexity measurement could not be reliably obtained due to syntactical or behavioral issues in the discovered model (i.e., a disconnected model or an unsound model). Additionally, to report the occurrence of a timeout or an exception during the execution of a discovery method we used ``\emph{t/o}'' and ``\emph{ex}'', respectively.

The first evaluation shows the absence of a clear winner among the discovery methods tested, although almost each of them clearly showed specific benefits and drawbacks.
%The first conclusion we can draw from the results of the default-parameters evaluation is that none of the discovery methods stood out as the best, although almost each of them clearly showed its benefits and drawbacks.

HILP experienced severe difficulties in producing useful outputs. The method often produced disconnected models or models containing multiple end places without providing information about the final marking (a well defined final marking is required in order to measure fitness and precision). Due to these difficulties, we could only assess model complexity for HILP, except for the simplest event log (the \emph{PRT5}), where HILP had performance comparable to the other methods.

$\alpha\$$ showed scalability issues, timing out in eight event logs (33\% of the times). Although none of the discovered models stood out in accuracy or in complexity, $\alpha\$$ in general produced models striking a good balance between fitness and precision (except for the \emph{BPIC13\textsubscript{inc}} log).

FO struggled in delivering sound models, discovering only eight sound models. Nevertheless, its outputs were usually highly fitting, scoring five times the best fitness.

S-HM\textsubscript{6} performed better than FO, although it also ended up producing unsound models. Of the 16 sound models discovered, nine scored the best fitness. However, precision varied according to the input event log, demonstrating that the performance of this method is bound to the type of log provided in input.%\newline

The remaining three methods, i.e., IM, ETM, and SM, consistently performed very well across the whole evaluation, excelling either in fitness, precision or f-score. IM scored 20 times a fitness greater than 0.90 (of which 8 times the highest), though, IM did not stand out for its precision. ETM and SM achieved 19 times a precision greater than 0.80, and ETM precision was the best 10 times. However, ETM scored high precision at the cost of lower fitness. Lastly, SM stood out for its F-score (i.e., high and balanced fitness and precision), achieving an F-score above 0.80 and outperforming the other methods for 18 times. Despite such good results, SM was not able to consistently produces sound models, as showed in the case of the PRT11 log.

In terms of complexity, IM, ETM, and SM delivered good results. IM and ETM always discovered sound and fully block-structured models (struct. of 1.00). ETM and SM discovered models that were within the two smallest models for more than the 80\% of the inputs, and that had low CFC, being the ones with the lowest CFC respectively on 14 and 6 logs. %specifically, ETM was able to discover 14 models with the lowest CFC, and SM 6 models.
On the execution time, SM was the clear winner. It systematically outperformed all the other methods, regardless the input. It was the fastest discovery method 23 times out of 24, discovering a model in less than a second over 19 logs. On the other hand, ETM was the slowest discovery method, reaching the timeout of four hours over 22 event logs.

Finally, Table~\ref{tab:hyperdata} displays the results of the hyper-parameter evaluation. The purpose of this second evaluation was to explore the solutions' space of each discovery method, to understand if they can achieve higher F-score when optimally tuned. According to this aim, the results of the hyper-parameter evaluation are extremely positive. All the discovery methods were able to improve their F-scores for almost all the inputs (values highlighted in italic in Table~\ref{tab:hyperdata}). Precisely, IM, FO, S-HM\textsubscript{6}, and SM achieved better F-scores in 19, 14, 15, and 11 event logs, respectively. This denotes that IM default parameters are not optimal when focusing on the discovery of models with high F-score. FO and S-HM\textsubscript{6} default parameters are more reliable. Whilst, SM showed to achieve the best with the default parameters more than the 50\% of the times. Lastly, SM again outperformed all the other methods, scoring the highest hyper-parameter optimized F-score over 20 event logs.

In conclusion, a method outperforming all other across all metrics could not be identified. All that aside, IM, ETM and SM showed to be the most effective methods when the focus is either on fitness, precision, or F-score, respectively. Nevertheless, all these three methods suffer from one common weakness, which is the inability to handle large-scale real-life logs, as reported for the PRT11 log in our evaluation. % reports for the case of the event log PRT11. Since, for this latter, we did not apply any pre-processing filtering (differently than the cases of BPIC14, BPIC15 set and BPIC17 event logs).
%In conclusion, despite there is not a best-for-all method, IM, ETM and SM definitely stood out respectively for fitness, precision, and F-score, performing well also in terms of complexity. Nevertheless, even such methods showed weaknesses. Among these, one common weakness was the inability to handle large-scale real-life logs, as our evaluation reports for the case of the event log PRT11. Since, for this latter, we did not apply any pre-processing filtering (differently than the cases of BPIC14, BPIC15 set and BPIC17 event logs). 
% !TEX root = main.tex
\section{Discussion}
\label{sec:discussion}

Our review highlights a growing interest in the field of automated process discovery,
and confirms the existence of a wide and heterogeneous number of proposals. 
Despite such a variety, we can clearly identify two main streams:
methods that output procedural process models, and methods that output declarative process models.
Further, while the latter ones only rely on declarative statements to represent a process,
the former provide various language alternatives, though, most of these methods output Petri nets.

The predominance of Petri nets is driven by the expressive power of this language,
and by the requirements of the methods used to assess the quality of the discovered process models (chiefly, fitness and precision).
%Indeed, as reported in Section~\ref{sec:benchmark}, quality assessment tools are based on Petri nets inputs.
Despite some modeling languages have a straightforward conversion to Petri nets,
the strict requirements of these quality assessment tools represent a limitation
for the proposals in this research field.
%Though this problem was already partially highlighted in the work of De Weerdt et al.~\cite{DeWeerdt}, at distance of five years it remains unsolved.
For the same reason, it was not possible to compare the two main streams,
so we decided to focus our evaluation and comparison on the procedural methods, which in any case,
have a higher practical relevance than their declarative counterparts,
given that declarative process models are hardly used in practice.
%since these latter have a higher relevance than the declarative ones.

   %% PROOFREADING STARTS HERE %%
Our benchmark shows benefits and drawbacks of the procedural automated process discovery methods, as well their limitations.
These latter include lack of scalability for large and complex logs, and strong differences in the output models, across the various quality metrics.
Regarding this aspect, the majority of methods were not able to excel in accuracy or complexity, except for IM, ETM and SM.
Indeed, these three ones were the only ones to consistently perform very well in fitness (IM), precision (ETM, SM), F-score (SM), complexity (IM, ETM, SM) and execution time (SM). Nevertheless, our evaluation shows that even IM, ETM and SM can fail when challenged with large-scale unfiltered real-life events logs, as shown in the case of the event log PRT11.
%In some cases (i.e., BPIC14, BPIC15, BPIC17 logs), in order to produce any output for which fitness and precision could be measured or to avoid extreme over-fitting (in the case of IM), we had to apply a noise-filtering technique as a pre-processing step.

To conclude, even if many proposals are available in this research area, and some of them are able to systematically deliver good to optimal results, there is still space for research and improvements. Furthermore, it is important to highlight that the great majority of the methods do not have a working or available implementation. This hampers their systematic evaluation, so one can only rely on the results reported in the respective papers. Finally, for those methods we assessed, we were not able to identify a unique winner, since the best methods
showed to either maximize fitness, precision or F-score. Despite these considerations, it can be noted that there has been significant progress in this field in the past five years. Indeed, IM, ETM and SM clearly outperformed the discovery methods developed in the previous decade and their extensions (i.e., Agnes and S-HM\textsubscript{6}).
   %% PROOFREADING ENDS HERE %%

% !TEX root = main.tex
\section{Threats to Validity}
\label{sec:threats_to_validity}

The first threat to validity refers to the potential selection bias and inaccuracies in data extraction and analysis typical of literature reviews. In order to minimize such issues, our systematic literature review carefully adheres to the guidelines outlined in \cite{Kitchenham}. Concretely, we used well-known literature sources and libraries in information technology to extract relevant works on the topic of automated process discovery. Further, we performed a backward reference search to avoid the exclusion of potentially relevant papers. Finally, to avoid that our review was threatened by insufficient reliability, we ensured that the search process could be replicated by other researchers. However, the search may produce different results as the algorithm used by source libraries to rank results based on relevance may be updated (see, e.g., Google Scholar).

% A second threat to validity derives from the fact that the search of relevant works on process discovery was performed in December 2016. This means that any work published after that date was not included in the benchmark. Moreover, only 32 primary studies were included in the SLR, with only 1-2 studies for each approach with information regarding research questions. This might endanger the overall accuracy of an approach's representation in the study.

% Finally, to avoid that our review was threatened by insufficient reliability, we ensured that the search process could be replicated by other researchers. However, the search may produce different results as internal processes of source libraries are frequently updated. %Additionally, since the process of creating a literature review also considers subjective factors
%(e.g., subjective interpretations considering inclusion criteria), other researchers may come to different conclusions and may not obtain exactly the same results as here presented.

The experimental evaluation on the other hand is limited in scope to techniques that produce Petri nets (or models in languages such as BPMN or Process Trees, which can be directly translated to Petri nets). Also, it only considers main studies identified in the SLR with an available implementation. %and takes the most recent incarnation of each method (e.g., we only considered S-HM\textsubscript{6}, as it was shown it outperforms the HM\textsubscript{6}). %When a more recent technique of a given type has been shown to outperform others in previous work, we took only the most recent one -- e.g., we only considered A\$, as it was shown in previous work to outperform the $\alpha$ and $\alpha+$ algorithms. Also, we used existing implementations of each retained technique with their default configuration. Since in every case the default configuration has been chosen by the authors of the approach, we have assumed that this configuration generally delivers better results.
In order to compensate for these shortcomings, we published the benchmarking toolset as open-source software in order to enable researchers both to reproduce the results herein reported and to run the same evaluation for other methods, or for alternative configurations of the evaluated methods.

Another limitation is the use of only 24 event logs, which to some extent limits the generalizability of the conclusions. However, the event logs included in the evaluation are all real-life logs of different sizes and features, including different application domains. To mitigate this limitation, we have structured the released benchmarking toolset in such a way that the benchmark can be seamlessly rerun with additional datasets.

% !TEX root = main.tex
\section{Related Work}
\label{sec:related_work}

A previous survey and benchmark of automated process discovery methods has been reported by De Weerdt et al.~\cite{DeWeerdt}. This survey covered 27 approaches, and it assessed 7 of them. We used it as starting point for our study.

%all of which are included in the 330 studies we identified during our systematic literature review (prior to filtering).

The benchmark reported by De Weerdt et al.~\cite{DeWeerdt} includes seven approaches, namely AGNEsMiner, $\alpha{+}$, $\alpha{++}$, Genetic Miner (and a variant thereof), Heuristics Miner and ILP Miner. In comparison, our benchmark includes $\alpha$\$ (which is an improved version of $\alpha{+}$ and $\alpha{++}$), Structured Heuristics miner (which is an extension of Heuristics Miner), Hybrid ILP Miner (an improvement of ILP), Evolutionary Tree Miner (which is a genetic algorithm postdating the evaluation of De Weerdt et al.~\cite{DeWeerdt})
%, which achieves better execution times by focusing on block-structured process models and simpler transformation rules.
Notably, we did not include AGNEsMiner due to the very long execution times (as suggested by the authors in a conversation over emails exchanged during this work).

Another difference with respect to the previous survey~\cite{DeWeerdt}, is that in our paper we based our evaluation both on public and proprietary event logs, whilst the evaluation of De Weerdt et al.~\cite{DeWeerdt} is solely based on artificial event logs and closed datasets, due to the unavailability of public datasets at the time of that study.

In terms of results, De Weerdt et al.\cite{DeWeerdt} found that Heuristics Miner achieved a better F-score than other approaches and generally produced simpler models, while ILP achieved the best fitness at the expense of low precision and high model complexity. Our results show that SM achieves even better F-score and lower model complexity than other techniques, followed by ETM and IM, which excelled for precision and fitness (respectively). Thus it appears that in this field, in the last years, progress has been pursued successfully.

Another previous survey in the field is outdated~\cite{vanderAalst2003} and a more recent one is not intended to be comprehensive~\cite{DBLP:conf/bpm/ClaesP12}, but rather limits on plug-ins available in the ProM toolset.
Another related effort is CoBeFra -- a tool suite for measuring fitness, precision and model complexity of automatically discovered process models~\cite{BrouckeWVB13}. %The methods for the evaluation of fitness and precision used in our benchmark are taken from this suite.

% and follow the recommendations in ....
%In addition, our benchmark includes approaches that postdate the evaluation reported in~\cite{DeWeerdt}. 
% !TEX root = main.tex
\section{Conclusion}
\label{sec:conclusion}

This article presented a Systematic Literature Review (SLR) of automated process discovery methods and a comparative evaluation of existing implementations of these methods using a benchmark covering twelve publicly-available real-life event logs, twelve proprietary real-life event logs, and nine quality metrics. The toolset used in this benchmark is available as open-source software and the 50\% of the event logs are publicly available. The benchmarking toolset has been designed in a way that it can be seamlessly extended with additional methods, event logs, and evaluation metrics.

%at \url{https://github.com/raffaeleconforti/ResearchCode} and all the event logs are sourced from the 4TU Centre and are available at: \url{https://data.4tu.nl/repository/collection:event_logs_real}.

The SLR put into evidence a vast number of automated process discovery methods (344 relevant papers were analysed). Traditionally, many of these proposals produce Petri nets, but more recently, we observe an increasing number of methods that produce models in other languages, including BPMN and declarative constraints. We also observe a recent emphasis on methods that produce block-structured process models.

%A large proportion of automated discovery methods take as point of departure the Directly-Follows Graph (DFG) and effectively apply heuristics over this graph to filter out infrequent behavior and to discover split and join gateways.  The heuristics miner is a key representative of this method but there are several other methods that rely on similar principles.

The results of the empirical evaluation show that methods that seek to produce block-structured process models (Inductive Miner and Evolutionary Tree Miner) achieve the best performance in terms of fitness or precision, and complexity. Whilst, methods that do not restrict the topology of the generated process models (Split Miner), produce process models of higher quality in terms of F-score, although these methods cannot guarantee soundness (though they can guarantee deadlock-freedom). We also observed that in the case of very complex event logs, it is necessary to use a filtering method prior to applying existing automated process discovery methods. Without this filtering, the precision of the resulting models was close to zero. A direction for future work is to develop automated process discovery techniques that incorporate adaptive filtering approaches so that they can auto-tune themselves to deal with very complex logs.

% None of the methods appears to have a sufficiently powerful and adaptive filtering technique to cope with the complexity of these event logs.
%The study has shown that there is still room for improvement in the field. Indeed, even the best performing methods showed to be unable to handle highly complex event logs in raw form. In order to handle the most complex logs, we had to apply a filtering technique prior to applying the process discovery algorithms in the benchmark.

Another limitation observed while conducting the benchmark, was the lack of universal measures of fitness and precision, which would be applicable not only to Petri nets (or BPMN models that can be mapped to Petri nets), but equally well to declarative or data-driven process modeling notations. Developing more universal measures of fitness and precision is another possible target of future work.

%One of the two top-performing automated process discovery methods (IM) is problematic because it often generates ``flower'' structures, i.e., fragments where a set of activities can be performed any number of times and in any order. These structures lead to low precision. The other top-performing method (ETM) suffers from significant performance issues, relying on a genetic algorithm at its core. Finally, S-HM\textsubscript{5.2}, a relatively well-performing method, is not robust -- it sometimes produces unsound process models due to its reliance on the Heuristics Miner.

% \appendices
% \input{appendix}

\section*{Acknowledgments}
This research is partly funded by the Australian Research Council (grant DP150103356) and the Estonian Research Council (grant IUT20-55). It is also partly supported by the H2020-RISE EU project FIRST (734599), the Sapienza grant DAKIP and the Italian projects Social Museum and Smart Tourism (CTN01\_00034\_23154), NEPTIS (PON03PE\_00214\_3), and RoMA - Resilence of Metropolitan Areas (SCN\_00064).

\newpage

\vspace{-2\baselineskip}
\begin{IEEEbiography}
[{\includegraphics[width=1in,height=1.25in,clip,keepaspectratio]{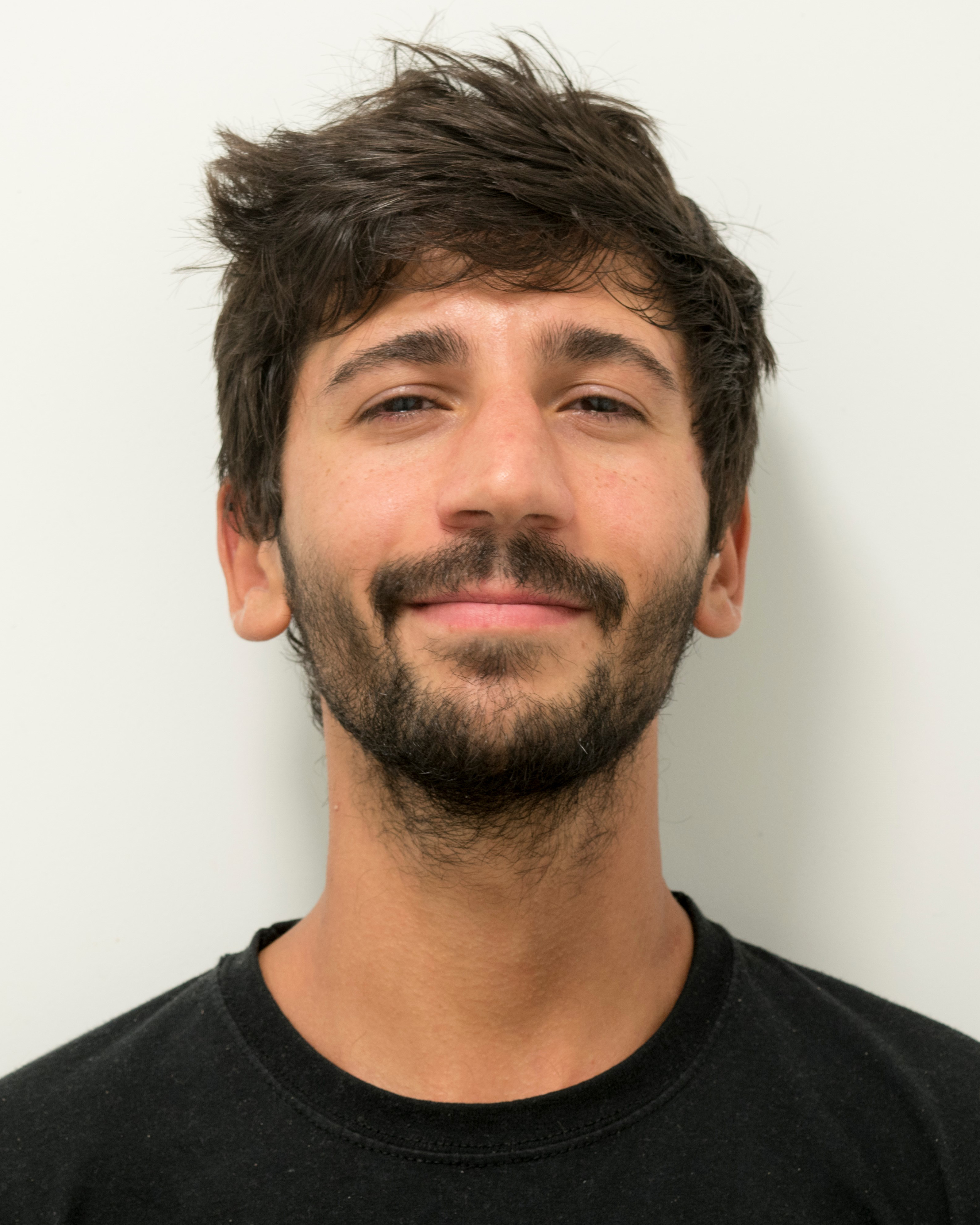}}]
{Adriano Augusto} is a joint-PhD student at University of Tartu (Estonia) and Queensland University of Technology (Australia). He graduated in Computer Engineering at Polytechnic of Turin (Italy) in 2016, presenting a master thesis in the field of Process Mining.
\end{IEEEbiography} 
\vspace{-2\baselineskip}
\begin{IEEEbiography}
  [{\includegraphics[width=25mm,height=32mm,clip,keepaspectratio]{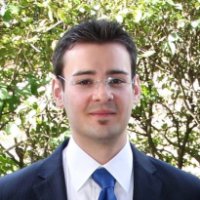}}]
  {Raffaele Conforti} is a Post-Doctoral Research Fellow at the Queensland University of Technology, Australia. He conducts research on process mining and automation, with a focus on automated process discovery, quality improvement of process event logs and process-risk management. %He obtained his PhD in Information Systems from Queensland University of Technology in 2014, his MSc in Computer Engineering from Universit\`{a} della Calabria in 2010, and his BSc in Computer Engineering from Universit\`{a} della Calabria in 2008. 
\end{IEEEbiography}
\vspace{-2\baselineskip}
\begin{IEEEbiography}
[{\includegraphics[width=1in,height=1.25in,clip,keepaspectratio]{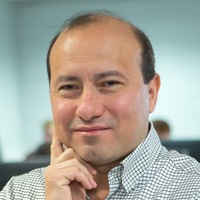}}]%
{Marlon Dumas} is Professor of Software Engineering at University of Tartu, Estonia. Prior to this appointment he was faculty member at Queensland University of Technology and visiting researcher at SAP Research, Australia. His research interests span across the fields of software engineering, information systems and business process management. He is co-author of the textbook ``Fundamentals of Business Process Management'' (Springer, 2013).
\end{IEEEbiography} 

\vspace{-2\baselineskip}
\begin{IEEEbiography}
  [{\includegraphics[width=25mm,height=32mm,clip,keepaspectratio]{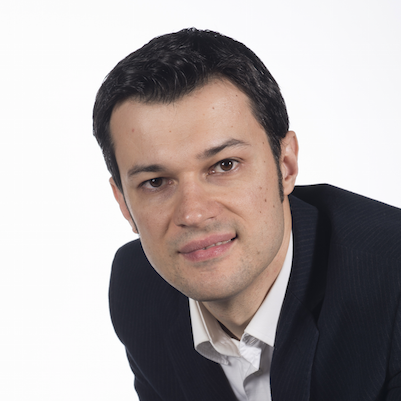}}]
  {Marcello La Rosa} is Professor of Information Systems at the Queensland University of Technology, Australia. His research interests include process mining, consolidation and automation. He leads the Apromore Initiative (www.apromore.org), a strategic collaboration between various universities for the development of a process analytics platform, and  co-authored the textbook ``Fundamentals of Business Process Management'' (Springer, 2013).
\end{IEEEbiography} 
\vspace{-2\baselineskip}
\vfill
\break \begin{IEEEbiography}%
[{\includegraphics[width=1in,height=1.25in,clip,keepaspectratio]{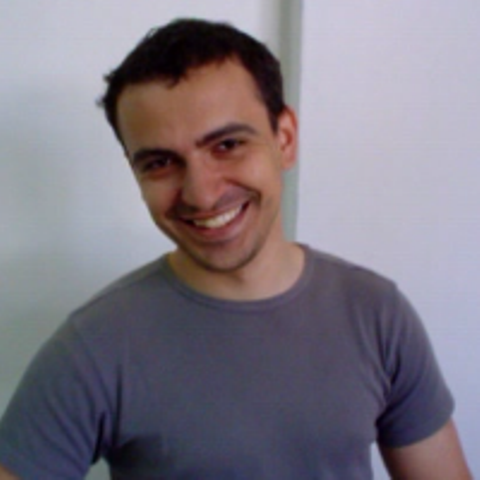}}]%
{Fabrizio Maria Maggi} is an Associate Professor at the University of Tartu, Estonia. He worked as Post-Doctoral Researcher at the  Department of Mathematics and Computer Science, Eindhoven University of Technology. His research interest span business process management, data mining and service-oriented computing.
\end{IEEEbiography} 
\vspace{-2\baselineskip}
\begin{IEEEbiography}[{\includegraphics[width=1in,height=1.25in,clip,keepaspectratio]{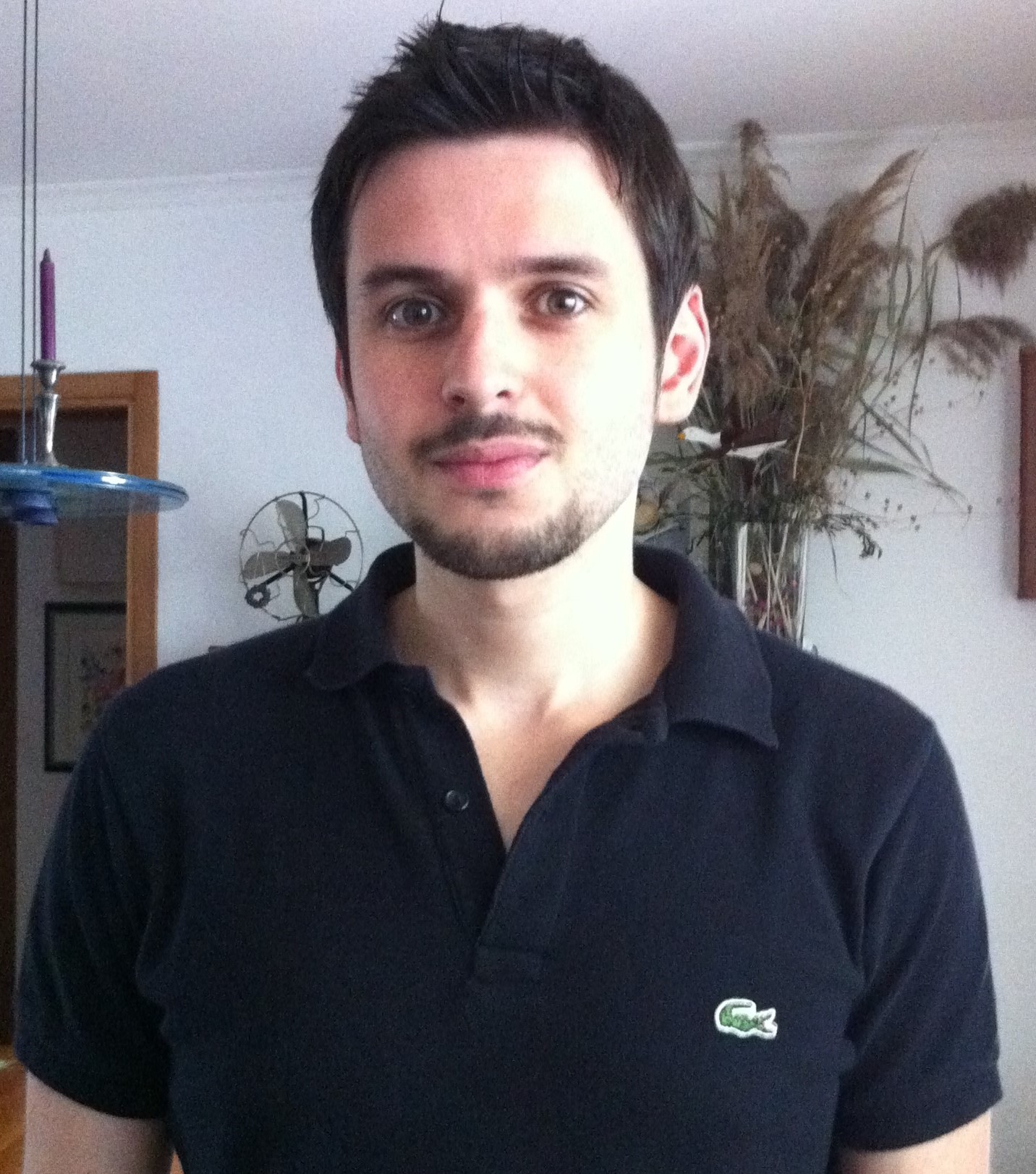}}]%
{Andrea Marrella}
is Post-Doctoral Research Fellow at Sapienza Universit\'a di Roma.
His research interests include human-computer interaction, user experience design, knowledge representation, reasoning about action, automated planning, business process management.
He has published over 40 research papers and articles and 1 book chapter on the above topics.
%, among others in ACM Transactions on Intelligent Systems and Technologies, IEEE Internet Computing and Journal on Data Semantics.
%%
%He obtained his PhD at Sapienza in 2013, with a thesis on the automated synthesis of recovery processes at run-time in case of unanticipated exceptions and exogenous events.
\end{IEEEbiography} 
\vspace{-2\baselineskip}
\begin{IEEEbiography}[{\includegraphics[width=1in,height=1.25in,clip,keepaspectratio]{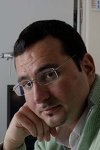}}]%
{Massimo Mecella} is Associate Professor (with a Full Professor Habilitation) with Sapienza Universit\`a di Roma. His research focuses on service oriented computing, business process management, Cyber-Physical Systems and Internet-of-Things, advanced interfaces and human-computer interaction, with applications in fields such as digital government, smart spaces, healthcare, disaster/crisis response \& management, cyber-security. 
He published more than 150 research papers and chaired different conferences in the above areas. %, and has a good expertise in research projects, especially in the European FP5, FP6, FP7 and Horizon2020.
\end{IEEEbiography} 

\vspace{-2\baselineskip}
\begin{IEEEbiography}[{\includegraphics[width=1in,height=1.25in,clip,keepaspectratio]{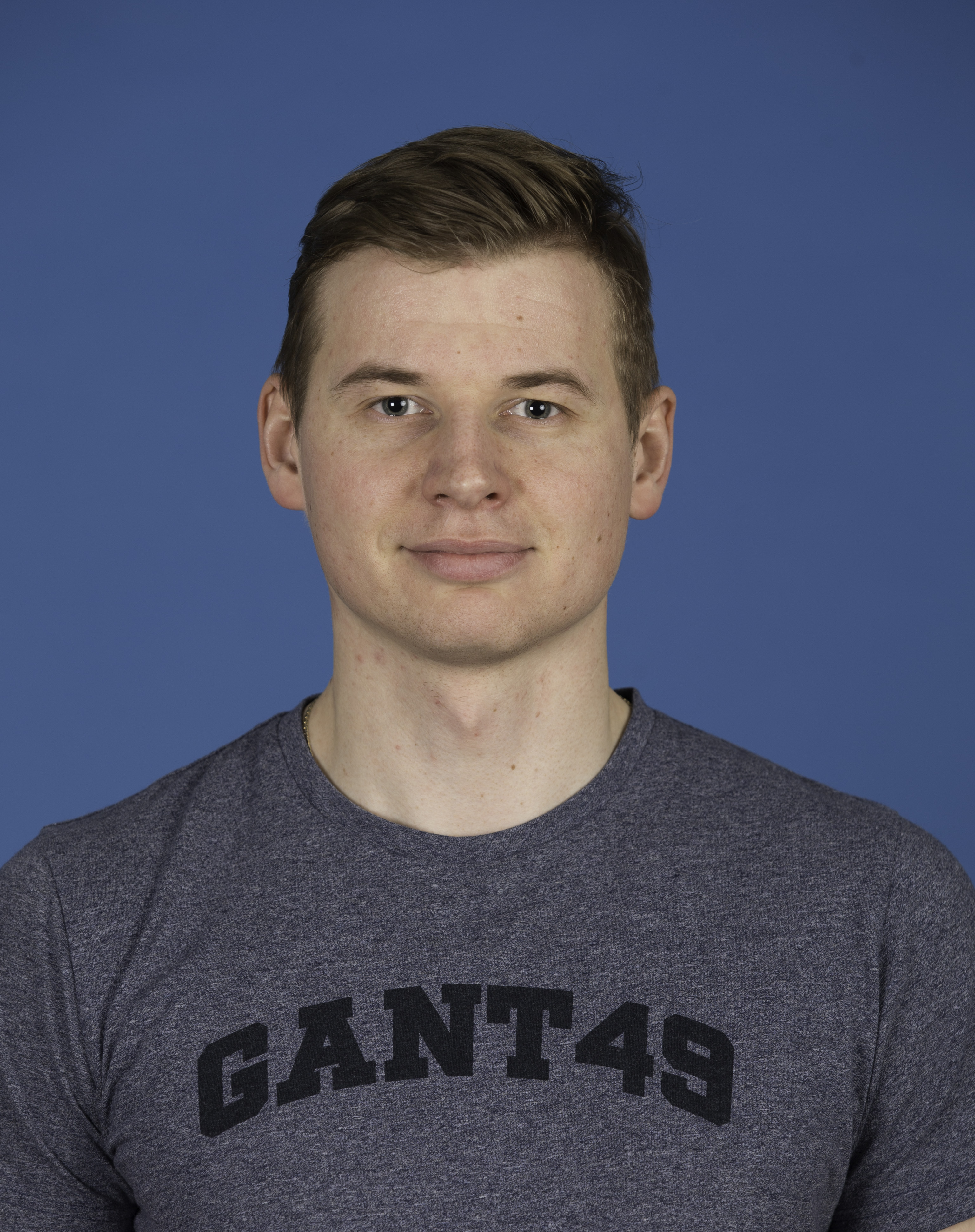}}]%
{Allar Soo}
Allar Soo is student in the Masters of Software Engineering at University of
Tartu. His Masters thesis is focused on automated process discovery and its use in practical settings.
\end{IEEEbiography} 
\vfill


%\vfill


\end{document}